\documentclass{elsart}
\usepackage{amssymb}

\usepackage{amsmath}

 \journal{European Journal of
Mechanics B/Fluids}
\input{tcilatex}

\begin{document}

\begin{frontmatter}


\title{Mixture of Fluids involving
Entropy Gradients\\ and Acceleration Waves in Interfacial Layers}
\author[Francia]{Henri Gouin\corauthref{cor}},
\ead{henri.gouin@univ.u-3mrs.fr}
\corauth[cor]{Corresponding Author}
%\thanksref{label3}
\author[Italia]{Tommaso Ruggeri}
\ead{ruggeri@ciram.unibo.it}
\ead[url]{http://www.ciram.unibo.it/ruggeri}

\address[Francia]{Laboratoire de Mod\'elisation en M\'ecanique et Thermodynamique  E.A. 2596\\
 Universit\'e Paul C\'ezanne-Aix-Marseille III, Case 322, 13397 Marseille Cedex 20 France}
\address[Italia]{Department of
Mathematics and Research Center of Applied Mathematics C.I.R.A.M.
University of Bologna Via Saragozza 8 40123-I Bologna Italy}

% use
%optional labels to link authors explicitly to addresses:
% \author[%
%label1,label2]{}

%%
\address{}

\begin{abstract}
% Text of abstract
Through an Hamiltonian action we write down the  system of equations
of motions for a mixture of thermocapillary fluids under the
assumption that the internal energy is a function not only of the
gradient of the densities but also of the gradient of the entropies
of each component. A Lagrangian associated with the kinetic energy
and the internal energy allows to obtain the equations of momentum
for each component and for the barycentric motion of the mixture. We
obtain also the balance of energy and we prove that the equations
are compatible with the second law of thermodynamics. Though the
system is of parabolic type, we prove that there exist two
tangential acceleration waves that characterize the interfacial
motion. The dependence of the internal energy of the entropy
gradients is mandatory for the existence of this kind of waves.
 The
differential system is non-linear but the waves propagate without
distortion due to the fact that they are linearly degenerate
(exceptional waves).

\end{abstract}

\begin{keyword}
Fluid mixtures, acceleration waves, interfacial layers

\PACS 47.55, 52.35, 68.05, 68.60

\end{keyword}

\end{frontmatter}

\section{Introduction}

Liquid-vapor and two-phase interfaces are generally represented by a
material surface endowed with an energy related to Laplace's surface
tension. In fluid mechanics and thermodynamics, the interface appears as a
surface separating two media.\ This surface has its own characteristic
behavior and energy properties \cite{Levitch}.\ Theoretical and experimental
detailed studies show that, when working far from critical conditions, the
capillary layer has a thickness equivalent to a few molecular beams \cite
{Ono}.\newline
Molecular models such as those used in kinetic theory of gas lead in
interfacial layers to laws of state associated with non-convex internal
energies, e.g., the van der Waals models \cite{cahn,rowlinson,Hohenberg}.
These models appear advantageous as they provide as even more precise
verification of Maxwell's rule applied to isothermal phase transition \cite
{gouin4}. Nonetheless, they present two disadvantages: \newline
First, for densities that lie between phase densities, the pressure may
become negative. Simple physical experiments can be used, however, to cause
traction that leads to these negative pressure values \cite{Rocard,Bruhat}.%
\newline
Second, in the field between bulks, internal energy cannot be represented by
a convex surface associated with the variation of densities and entropy.
This fact seems to contradict the existence of steady equilibrium state of
the matter in this type of region. \newline
To overcome these disadvantages, the thermodynamic investigative replaces
the non-convex portion corresponding to internal energy with a plane domain.
The fluid can no longer be considered as a continuous medium. The interface
is represented as a material surface with a null thickness. In this case,
the only possible representation of the dynamic behavior of the interface is
one of a discontinuous surface, and its essential structure remains unknown.

In the equilibrium state it is possible to eliminate the above disadvantages
by appropriately modifying the stress tensor of the capillary layer, which
is expressed in an anisotropic form. As a consequence, the energy of the
continuous medium must change \cite{cahn,rowlinson,Bongiorno}. A
representation of the energy near the critical point therefore allows the
study of interfaces of non-molecular size. This approach is not new and, in
fact, dates back to van der Walls \cite{vdW} and Korteweg \cite{Korteweg};
it corresponds to what is known as the Landau-Ginzburg theory \cite
{Hohenberg}. The representation proposed in the present study is based on
the notion of internal energy which is more convenient to use when the
temperature is not uniform. \ One of the problems that complicates this
study of phase transformation dynamics is the apparent contradiction between
Korteweg's classical stress theory and the Clausius-Duhem inequality \cite
{Gurtin}. Proposal made by Eglit \cite{Eglit}, Dunn and Serrin \cite{Dunn},
Casal and Gouin \cite{Gouin5} and others rectify this apparent anomaly for
liquid-vapor interfaces of a pure fluid.

To study capillary layers and bulk phases, the simplest model in continuum
mechanics considers an internal energy $\varepsilon$ as the sum of two
terms: a first one corresponding to a medium with a uniform composition
equal to the local one and a second one associated with the non-uniformity
of the fluid \cite{cahn,vdW}. The second term is approximated by a gradient
expansion, typically truncated to the second order. In the simplest version
of the theory we have $\varepsilon =\rho \,\alpha (\rho ,s)+{\frac{1}{2}}%
\,m\,(\mathbf{\nabla }\rho )^{2}$, where $\rho $ is the matter density, $s$
the specific entropy, $\alpha $ the specific internal energy of the fluid
assumed to be homogeneous and $m$ is a coefficient independent of $s$, $\mathbf{%
\nabla }\rho $ and of any higher derivatives. Obviously, the model
is simpler than models associated with the renormalization-group
theory \cite {domb}. Nevertheless, it has the advantage of easily
extending well-known results for equilibrium cases to the dynamics
of interfaces \cite {slemrod,trusk}. In such a model, $s$ varied
with $\rho $ through the interface in the same way as in the bulk,
then $s$ would always be that function of $\rho $ which, at given
temperature $T$ satisfied
\begin{equation}
\frac{\partial \alpha }{\partial s}(\rho ,s)=T.  \label{temp0}
\end{equation}
With this assumption, $s=s(\rho )$ and $\varepsilon =\varepsilon (\rho )$,
so that the points representing single-phase states in the $\rho
,s,\varepsilon $ space lie on a curve instead of on a surface $\varepsilon
=\varepsilon (\rho ,s)$. This was the original assumption of van der Waals
which was later justified by Ornstein in 1909 for a system composed of
molecules with long-ranged but weak attractive force; \textit{this
assumption is not exact for more realistic potentials}\footnote{%
Main sentences in this paragraph and more precisely the comments on
static interfaces are issued from the book of Rowlinson and Widom
\cite{rowlinson} and its bibliography.}.\newline As coexistence
curves these are no way peculiar; the only peculiarity is that the
single-phase states -in this version of this approximation- have
collapsed onto the coexistence curve instead being represented by
the points of an extended two-dimensional region of which the
coexistence curve is merely a boundary. There is then no proper
two-density description of the one-phase states of a one-component
system in the lowest order of the mean-field approximation. There is
such a description of the two-states, where not even in mean-field
approximation is there any discernible peculiarity; but in practice
the potential for the two-density form of the van der Waals theory
is then not constructed by the prescription in (\ref{temp0}) but by
other means. For example Rowlinson and Widom introduce in
\cite{rowlinson}, chapter 9, an energy arising from the mean-field
theory and depending on the deviations of the densities $s$ and
$\rho $, say, from their values at the critical point and the
gradients of these densities. It is also seen that in $c$-components
systems, $c+1$ densities -the densities $\rho _{1},...,\rho _{c}$ of
the $c$ components and the entropy density $s$- may vary
independently through the interface.\newline Aside from the question
of accuracy, there are also qualitative features of some interfaces
in physical-chemistry, especially in systems of more than one
component, that require two or more independently varying densities
for their description. An example is strong positive or negative
adsorption of a component $i\mathit{\ }$associated with a
non-monotonic profile $\rho _{i}(z) $ where $z$ is the spatial
variable. In the one-density theory based
on the approximation $\frac{\partial \alpha }{\partial s}-T=0\ $and $\frac{%
\partial \alpha }{\partial \rho _{i}}=0$ for all $j\neq i$, the resulting
one-density model leads inevitably to a monotonic $\rho _{i}(z).$ In
a theory based on two or more densities, by contrast, we may have a
realistic trajectory with which is associated non-monotonic behavior
of one or more of the components if we suppose $\rho _{j}(z)$ to be
monotonic. \newline We must also allow the independent variation of
at least two densities, entropy included, if we are to account
contact angles in three-phase equilibrium : J. Cahn made the remark
that we might use a two- (or more-) density van der Waals theory to
describe the case of non-spreading in the three phase equilibrium, (\cite{rowlinson}%
, chapter 8). Then, at equilibrium, Rowlinson and Widom pointed out
that for single fluids the model must be extended by taking into
account not only the strong variations of matter density through the
interfacial layer but also the strong variations of entropy.\newline
Also in dynamics, for an \textit{extended Cahn and Hilliard fluid},
the volume internal energy $\varepsilon $ is proposed with a
gradient expansion depending not only on $\mathrm{grad}\ \rho \,$
but also on $\mathrm{grad}\ s$ corresponding to a strong heat supply
in the change of phases: $\varepsilon =f(\rho ,s,\mathbf{\nabla
}\rho ,\mathbf{\nabla }s).$ The medium is then called a
\textit{thermocapillary fluid} \cite{casal,Gouin6}.

To extend the model to fluid mixtures corresponding to $c$-component
systems and realistic potentials in molecular theory of fluid
interfaces, the internal energy is assumed to be a functional of the
different densities of the mixture. In all the cases where strong
gradient of densities occurs - for example shocks or capillary
layers - the internal energy is chosen as a function of successive
derivatives of densities of matter and entropies. To be in
accordance with the physical phenomena presented in
\cite{rowlinson}, one will consider the internal energy of a
two-component fluid mixture with an internal energy depending also
on the gradients of entropy of each component. The internal energy
is a Galilean invariant, it does not depend on the reference frame;
hence the internal energy depends also on the relative velocity
between the two components of the mixture.\newline The conservative
motions of \textit{thermocapillary fluid mixture} are relevant to
the so-called second gradient theory \cite{Germain} and we obtain a
complete set of balance equations for conservative motions; we
extend this result to the dissipative case. Our goal is simply to
verify the consistency of our model with the second law of
thermodynamics. We consider a special case of dissipative
thermocapillary mixtures where the introduction of dissipative
forces is only done in the framework of the first gradient theory;
we deduce the Fick law as a consequence of a friction behavior
between the components and from the equations of motion of the
components. In such a case, extended thermodynamic principle (as
Gibbs identity) provides a set of equations that satisfy the entropy
principle, thereby making these irreversible motions compatible with
the second law of thermodynamics.\newline

The idea of studying interface motions as localized traveling waves
in a multi-gradient theory is not new and can be traced throughout
many problems of condensed matter and phase-transition physics
\cite{gouin2}. In Cahn and Hilliard's model \cite{cahn}, the
direction of solitary waves was along the gradient of density
\cite{slemrod,gouin2}. The introduction of the model of
thermocapillary fluid mixture provides a better understanding of the
behavior of motions in fluid mixture interfaces: it is possible to
obtain the previous solitary waves but also a new kind of adiabatic
waves may be forecasted. These waves are associated with the spatial
second derivatives of entropy and matter densities. For this new
kind of adiabatic waves, the direction of propagation is normal to
the gradient of densities. In the case of a thick interface, the
waves are tangential to the interface and the wave velocities depend
on the constitutive equations. Finally we observe that, also if the
differential system associated with the wave motions is non-linear,
the waves propagate without distortion due to the fact that they are
linearly degenerate (exceptional waves)\cite{Ruggeri}.

\section{Equations of thermocapillary mixtures}

\subsection{Conservative motions of thermocapillary mixtures}

To derive the governing equations and boundary conditions in the
dissipative-free case, we use the Hamilton principle of least action \cite
{Gavrilyuk 2}. In continuum mechanics, the principle with a system endowed
with an infinite number of degrees of freedom was initiated by Lin \cite{Lin}%
, Herivel \cite{Herivel}, Serrin \cite{Serrin}, Berdichevsky \cite
{Berdichevsky} and many others; it was proposed by Gouin for fluid mixtures
\cite{gouin3}.\ The main idea is to propose a Lagrangian which yields the
behavior of the medium as the difference between a kinetic and a potential
energy per unit volume.\ Then, the variations of the Hamilton action
obtained as a linear functional of virtual displacements allow to find the
governing equations and boundary conditions. For real media, the
irreversibility is introduced both in equations of motion and equation of
energy by using a classical approach through the dissipative function,
diffusion and heat fluxes.

We study a mixture of two fluids: the motion of a two-fluid continuum can be
represented using two surjective mappings (i=1,2)
\begin{equation*}
(t,\mathbf{x)\rightarrow X}_{i}=\mathbf{\Lambda }_{i}(t,\mathbf{x)}
\end{equation*}
where $(t,\mathbf{x)}$ belongs to $\left[ t_{1},t_{2}\right] \times D_{t}$,
a set in the time-space occupied by the fluid between times $t_{1}$ and $%
t_{2}$. Variables $\mathbf{X}_{i}$ denote the positions of each component of
the mixture in reference spaces $D_{0i}$. Variations of particle motions are
deduced from virtual motions
\begin{equation*}
\mathbf{X}_{i}=\mathbf{\Xi }_{i}(t,\mathbf{x,}\varkappa _{i}\mathbf{)},
\end{equation*}
where scalars $\varkappa _{i}$ are defined in a neighborhood of zero; they
are associated with a two parameter family of virtual motions of the
mixture. The real motion corresponds to $\varkappa _{i}=0$, the associated
virtual displacements are \cite{gouin3}
\begin{equation*}
\delta _{i}\mathbf{X}_{i}=\frac{\partial \mathbf{\Xi }_{i}}{\partial
\varkappa _{i}}\left\vert _{\varkappa _{i}=0}\right. .
\end{equation*}
They generalize what is obtained for a single fluid \cite{gouin0}. To the
virtual displacements\ $\delta _{i}\mathbf{X}_{i},$ we associated its image$%
\mathbf{\ \zeta }_{i}$ in the physical space$\ D_{t}$ occupied by the fluid
mixture at time $t\ $\cite{Gavrilyuk 2,Gouin 6},
\begin{equation*}
\mathbf{\zeta }_{i}=-\frac{\partial \mathbf{x}}{\partial \mathbf{X}_{i}}%
\delta _{i}\mathbf{X}_{i}.
\end{equation*}
Conservation of matter for each component requires that
\begin{equation}
\rho _{i}\text{ det}\left( \frac{\partial \mathbf{x}}{\partial \mathbf{X}_{i}%
}\right) = \rho _{io}\left( \mathbf{X}_{i}\right) ,  \label{mass0}
\end{equation}
where $\rho _{io}$ is the reference volume mass in $D_{0i}$ and $\text{ det}%
\left( \frac{\partial \mathbf{x}}{\partial \mathbf{X}_{i}}\right)$ the
Jacobian determinant of the motion of component $i$. In differentiable cases
eqs (\ref{mass0}) are equivalent to the equations of balance of matter
densities $\rho _{i}$
\begin{equation}
{\frac{\partial \rho _{i}}{\partial t}}\ +\ \text{div}(\rho _{i}{\mathbf{u}}%
_{i})=0,  \label{mass i}
\end{equation}
where ${\mathbf{u}}_{i}$ denotes the velocity vectors of each component $i$.
Now, we assume that the mixture has an entropy for each component \cite{ET};
for conservative motions, the equations of conservation of specific
entropies $s_{i}$ are
\begin{equation}
{\frac{\partial \rho _{i}s_{i}}{\partial t}}\ +\ \mathrm{div}(\rho _{i}s_{i}{%
\mathbf{u}}_{i})=0.  \label{entropy i}
\end{equation}
Then, relations
\begin{equation*}
s_{i}=s_{io}\left( \mathbf{X}_{i}\right)
\end{equation*}
define an isentropic motion of the fluid mixture. We deduce the following
relations of tensorial quantities \cite{Gavrilyuk 2,gouin3}
\begin{equation}
\delta _{i}\mathbf{u}_{i}=\frac{d_{i}\mathbf{\zeta }_{i}}{dt}-\frac{\partial
\mathbf{u}_{i}}{\partial \mathbf{x}}\mathbf{\zeta }_{i},\ \ \delta _{i}\rho
_{i}=-\mathrm{div}\ (\rho _{i}\mathbf{\zeta }_{i}),\ \ \delta _{i}s_{i}=-%
\frac{\partial s_{i}}{\partial \mathbf{x}}\mathbf{\zeta }_{i}.  \label{App1}
\end{equation}
where $\dfrac{d_{i}}{dt}=\dfrac{\partial }{\partial t}+{\mathbf{u}}_{i}.%
\mathbf{\nabla }$ denotes the material derivative relatively to the
component $i$. We assume that the volume potential energy of the
mixture is in the form
\begin{equation*}
\varepsilon =\epsilon (\rho _{i},s_{i},{\mathbf{\nabla }}\rho _{i},{\mathbf{%
\nabla }}s_{i},{\mathbf{w}}),
\end{equation*}
with ${\mathbf{w}}={\mathbf{u}}_{1}-{\mathbf{u}}_{2}$ the relative velocity
of the two components of the mixture. This means that the fluid mixture is a
function not only of the densities of matter $\rho _{i}\ $and specific
entropies $s_{i}\ $but also of the gradients of $\rho _{i}$ and $s_{i}$. The
fact that $\varepsilon $ depends on two entropies is classically adopted in
the literature \cite{gouin3,bedford,Muller}. Moreover, for a two-velocity
medium, there is no coordinate system within the framework of which any
motion could be disregarded.\ So, the standard definition of potential
energy leads to its dependence on the relative motion of the components. The
dependence of $\varepsilon$ with respect to the relative velocity is analog
to take into account the added mass effect in heterogeneous two-fluid theory
as done by berdichevsky \cite{Berdichevsky} and Geurst \cite{geurst1,geurst2}%
. Let us note we can assume also that $\varepsilon $ is depending on $(t,%
\mathbf{x)}$; by this way, we introduce directly the extraneous potential of
the body forces.\newline
The potential $\varepsilon$ is related with the volume internal energy $%
\varpi$ of the mixture through the transformation
\begin{equation*}
\varpi = \varepsilon - \frac{\partial\varepsilon}{\partial{\mathbf{w}}}\, {%
\mathbf{w}},
\end{equation*}
so that,
\begin{equation*}
e = \sum_{i=1}^{2}\frac{1}{2}\rho _{i}\mathbf{u}_{i}^{2}+\varpi
\end{equation*}
is the total energy of the system \cite{Gavrilyuk 2,Gavrilyuk 3}.\newline
The equation of motion of component $i\,\ $is given by a variational method
associated with a Hamilton action; the vector field $\mathbf{x\in }$ $D_{t}\
\mathbf{\rightarrow \ \zeta }_{i}$ is two time continuously differentiable.
The Lagrangian of the mixture is
\begin{equation*}
L=\sum_{i=1}^{2}\frac{1}{2}\rho _{i}\mathbf{u}_{i}^{2}-\varepsilon ,
\end{equation*}
and consequently, the Hamilton action between the times $t_{1}$ and $t_{2}$
is
\begin{equation*}
I=\int_{t_{1}}^{t_{2}}\int_{D_{t}}L\ d\mathbf{x}dt.
\end{equation*}
From the definition of virtual motions, we obtain immediately two variations
of the action of Hamilton associated with $i=1,2$,
\begin{eqnarray*}
\delta _{i}I &=&\int_{t_{1}}^{t_{2}}\int_{D_{t}}\left( \left( \frac{1}{2}{u}%
_{i}^{2}-\epsilon _{,\rho _{i}}\right) \ \delta _{i}\rho _{i}-\epsilon
_{,\rho _{i,\gamma }}\ \delta _{i}\rho _{i,\gamma }+\rho _{i}{K}_{i\gamma }\
\delta _{i}{u}_{i\gamma }\right. \\
&&\left. -\epsilon ,_{s_{i}}\delta _{i}s_{i}\ -\epsilon _{,s_{i,\gamma
}}\delta _{i}s_{i,\gamma }\right) \ d\mathbf{x}dt\ ,
\end{eqnarray*}
where subscript $\gamma $ corresponds to spatial derivatives associated with
gradient terms; as usually summation is made on repeated subscript $\gamma$ from $1$ to $3$; $\mathbf{K}_{i}=\mathbf{%
u}_{i}+(-1)^{i}\dfrac{1}{\rho _{i}}\left( \dfrac{\partial \epsilon }{%
\partial \mathbf{w}}\right) ^{T}$ where index $^{T}$ denotes the
transposition. Then, by integration by part we obtain,
\begin{eqnarray*}
\delta _{i}I &=&\int_{t_{1}}^{t_{2}}\int_{D_{t}}\left( \left( \frac{1}{2}{u}%
_{i}^{2}-\epsilon _{,\rho _{i}}+(\epsilon _{,\rho _{i,\gamma }})_{,\gamma
}\right) \ \delta _{i}\rho _{i}+\rho _{i}{K}_{i\gamma }\ \delta _{i}{u}%
_{i\gamma }\right. \\
&&-\left. \left( \epsilon ,_{s_{i}}-(\epsilon _{,s_{i,\gamma }})_{,\gamma
}\right) \ \delta _{i}s_{i}-(\epsilon _{,\rho _{i,\gamma }}\delta _{i}\rho
_{i})_{,\gamma }-(\epsilon _{,s_{i,\gamma }}\delta _{i}s_{i})_{,\gamma
}\right) d\mathbf{x}dt.
\end{eqnarray*}
Let us denote by
\begin{equation}
\rho _{i}\theta _{i}\ \equiv \ \frac{\widehat{\partial }\epsilon }{\widehat{%
\partial }s_{i}}\mathrm{\ \ \ and}\ \ \ h_{i}\equiv \frac{\widehat{\partial }%
\epsilon }{\widehat{\partial }\rho _{i}},  \label{temperature&enthalpie}
\end{equation}
where $\widehat{\partial }$ is the \textit{variational derivative operator.}
That is to say,
\begin{equation*}
\rho _{i}\theta _{i}\ =\epsilon ,_{s_{i}}-(\epsilon _{,s_{i,\gamma
}})_{,\gamma }\equiv \epsilon _{,s_{i}}-\mathrm{div\ }\mathbf{\Psi }_{i}%
\mathrm{\ \ and}\ \ h_{i}=\epsilon _{,\rho _{i}}-(\epsilon _{,\rho
_{i,\gamma }})_{,\gamma }\equiv \epsilon _{,\rho _{i}}-\mathrm{div\ }\mathbf{%
\Phi }_{i}\mathrm{\ },
\end{equation*}
with,
\begin{equation*}
\mathbf{\Psi }_{i}\equiv \frac{\partial \epsilon }{\partial \mathbf{\nabla }%
s_{i}}\mathrm{\ \ and}\ \ \mathbf{\Phi }_{i}\equiv \frac{\partial \epsilon }{%
\partial \mathbf{\nabla }\rho _{i}}.
\end{equation*}
Introducing $R_{i}=\frac{1}{2}u_{i}^{2}-h_{i},$ and taking into account of
the expressions for $\theta _{i}$ and $h_{i}$ given by eqs (\ref
{temperature&enthalpie}), we get
\begin{equation*}
\delta _{i}I=\int_{t_{1}}^{t_{2}}\int_{D_{t}}\left\{ R_{i}\ \delta _{i}\rho
_{i}+\rho _{i}\mathbf{K}_{i}\cdot \delta _{i}\mathbf{u}_{i}-\rho _{i}\theta
_{i}\ \delta _{i}s_{i}-\mathrm{div}\left( \mathbf{\Phi }_{i}\ \delta
_{i}\rho _{i}+\mathbf{\Psi }_{i\ }\delta _{i}s_{i}\right) \right\} d\mathbf{x%
}dt,
\end{equation*}
and from relations (\ref{App1}), we obtain
\begin{eqnarray*}
\delta _{i}I &=&\int_{t_{1}}^{t_{2}}\int_{D_{t}}\left\{ -R_{i}\ \mathrm{div}%
\ (\rho _{i}\mathbf{\zeta }_{i})+\rho _{i}\mathbf{K}_{i}\cdot \left( \frac{%
d_{i}\mathbf{\zeta }_{i}}{dt}-\frac{\partial \mathbf{u}_{i}}{\partial
\mathbf{x}}\mathbf{\zeta }_{i}\right) \right. \\
&&\left. +\,\rho _{i}\theta _{i}\ \frac{\partial s_{i}}{\partial \mathbf{x}}%
\,\mathbf{\zeta }_{i}-\mathrm{div}\left( \mathbf{\Phi }_{i}\ \delta _{i}\rho
_{i}+\mathbf{\Psi }_{i}\ \delta _{i}s_{i}\right) \right\} d\mathbf{x}dt.
\end{eqnarray*}
Consequently,
\begin{eqnarray*}
\delta _{i}I &=&\int_{t_{1}}^{t_{2}}\int_{D_{t}}\left\{ \rho _{i}\left(
\frac{\partial R_{i}}{\partial \mathbf{x}}+\theta _{i}\frac{\partial s_{i}}{%
\partial \mathbf{x}}-\frac{d_{i}\mathbf{K}_{i}^{T}}{dt}\ -\mathbf{K}_{i}^{T}%
\frac{\partial \mathbf{u}_{i}}{\partial \mathbf{x}}\right) \ \mathbf{\zeta }%
_{i}\right. \\
&& \\
&&+\frac{\partial }{\partial t}(\rho _{i}\mathbf{K}_{i}^{T}\ \mathbf{\zeta }%
_{i})-\mathrm{div}\left( \rho _{i}R_{i}\ \mathbf{\zeta }_{i}-\rho _{i}%
\mathbf{u}_{i}\ \mathbf{K}_{i}^{T}\ \mathbf{\zeta }_{i}+\mathbf{\Phi }_{i}\
\delta _{i}\rho _{i}+\mathbf{\Psi }_{i}\ \delta _{i}s_{i}\right) \Big\}\;d%
\mathbf{x}dt.
\end{eqnarray*}
The Stokes formula and relation (\ref{App1}) yield
\begin{eqnarray}
&&\delta _{i}I=\int_{t_{1}}^{t_{2}}\int_{D_{t}}\rho _{i}\left( \frac{%
\partial R_{i}}{\partial \mathbf{x}}+\theta _{i}\frac{\partial s_{i}}{%
\partial \mathbf{x}}-\frac{d_{i}\mathbf{K}_{i}^{T}}{dt}\ -\mathbf{K}_{i}^{T}%
\frac{\partial \mathbf{u}_{i}}{\partial \mathbf{x}}\right) \ \mathbf{\zeta }%
_{i}\ d\mathbf{xdt}  \label{Hamilton variations} \\
&&\ \ \ \ \ \ \ \ \ \ \ \ \ \ \ \ \ \ \ \ \ \ \ \ \ \ \ \ \mathbf{+}%
\int_{t_{1}}^{t_{2}}\int_{\partial D_{t}}\ \ \ g\ \rho _{i}\mathbf{K}%
_{i}^{T}\ \mathbf{\zeta }_{i}  \notag \\
&&-\mathbf{n}.\left( \rho _{i}R_{i}\ \mathbf{\zeta }_{i}-\rho _{i}\mathbf{u}%
_{i}\ \mathbf{K}_{i}^{T}\ \mathbf{\zeta }_{i}-\mathbf{\Phi }_{i}\ \frac{%
\partial \rho _{i}}{\partial \mathbf{x}}\ \mathbf{\zeta }_{i}-\mathbf{\Psi }%
_{i}\ \frac{\partial s_{i}}{\partial \mathbf{x}}\;\mathbf{\zeta }_{i}-\rho
_{i}\mathbf{\Phi }_{i\ }\mathrm{div\ }\mathbf{\zeta }_{i}\right) d\sigma _{%
\mathbf{x}}dt,  \notag
\end{eqnarray}
\newline
where $\partial D_{t}$ (of mesure $d\sigma _{\mathbf{x}}$) is the boundary
of $D_{t}$, $\mathbf{n}$ is the unit external normal vector to $\partial
D_{t}$\ and $g$ is the velocity of $\partial D_{t}$. If we consider a vector
field $\mathbf{x\in }$ $D_{t}\ \mathbf{\rightarrow \ \zeta }_{i}$ and its
first derivatives vanishing simultaneously on the boundary $\partial D_{t},$
the Hamilton principle expressed in the form: $\forall \mathbf{\ \zeta }%
_{i},\ \ \delta _{i}a=0\,$ leads to
\begin{equation*}
\forall \mathbf{\ \zeta }_{i},\ \ \int_{t_{1}}^{t_{2}}\int_{D_{t}}\rho
_{i}\left( \frac{\partial R_{i}}{\partial \mathbf{x}}+\theta _{i}\frac{%
\partial s_{i}}{\partial \mathbf{x}}-\frac{d_{i}\mathbf{K}_{i}^{T}}{dt}\ -%
\mathbf{K}_{i}^{T}\frac{\partial \mathbf{u}_{i}}{\partial \mathbf{x}}\right)
\ \mathbf{\zeta }_{i}\ d\mathbf{x}dt\ =0
\end{equation*}
and consequently,
\begin{equation}
\frac{d_{i}\mathbf{K}_{i}}{dt}+\left( \frac{\partial \mathbf{u}_{i}}{%
\partial \mathbf{x}}\right) ^{T}\mathbf{K}_{i}=\mathbf{\nabla }R_{i}+\theta
_{i}\mathbf{\nabla }s_{i}\ .  \label{App. 2}
\end{equation}
Let us note that the value of the first member of eq. (\ref{App. 2}) is
equal to
\begin{equation*}
\frac{d_{i}\mathbf{u}_{i}}{dt}+(-1)^{i}\frac{d_{i}}{dt}\left( \frac{1}{\rho
_{i}}{\frac{\partial \epsilon }{\partial {\mathbf{w}}}}\right) ^{T}+\left(
\frac{\partial \mathbf{u}_{i}}{\partial \mathbf{x}}\right) ^{T}\mathbf{u}%
_{i}+\frac{(-1)^{i}}{\rho _{i}}\left( \frac{\partial \mathbf{u}_{i}}{%
\partial \mathbf{x}}\right) ^{T}\left( {\frac{\partial \epsilon }{\partial {%
\mathbf{w}}}}\right) ^{T}
\end{equation*}
and eq. (\ref{App. 2}) yields
\begin{eqnarray*}
&&\rho _{i}\frac{d_{i}\mathbf{u}_{i}}{dt}+(-1)^{i}\ \left( \mathrm{div\;}{%
\mathbf{u}}_{i}\ \left( {\frac{\partial \epsilon }{\partial {\mathbf{w}}}}%
\right) ^{T}+\frac{d_{i}}{dt}\left( {\frac{\partial \epsilon }{\partial {%
\mathbf{w}}}}\right) ^{T}+\left( \frac{\partial \mathbf{u}_{i}}{\partial
\mathbf{x}}\right) ^{T}\left( {\frac{\partial \epsilon }{\partial {\mathbf{w}%
}}}\right) ^{T}\right) \\
&&=\rho _{i}\theta _{i}\,{\mathbf{\nabla }}\,s_{i}-\rho _{i}\,{\mathbf{%
\nabla }}\,h_{i}\ .
\end{eqnarray*}
Taking into account of eq. (\ref{mass i}) of conservation of mass of component $%
i,$ we deduce
\begin{eqnarray*}
&&\frac{\partial {\rho _{i}\mathbf{u}}_{i}}{\partial t}\,+\,\mathrm{div}%
(\rho _{i}{\mathbf{u}}_{i}\otimes {\mathbf{u}}_{i})+(-1)^{i}\left( \mathrm{%
div\;}{\mathbf{u}}_{i}\ \left( {\frac{\partial \epsilon }{\partial {\mathbf{w%
}}}}\right) ^{T}+\frac{\partial }{\partial t}\left( {\frac{\partial \epsilon
}{\partial {\mathbf{w}}}}\right) ^{T}\right. \\
&&\left. +\frac{\partial }{\partial \mathbf{x}}\left( {\frac{\partial
\epsilon }{\partial {\mathbf{w}}}}\right) ^{T}{\mathbf{u}}_{i}+\left( \frac{%
\partial {\mathbf{u}}_{i}}{\partial {\mathbf{x}}}\right) ^{T}\left( {\frac{%
\partial \epsilon }{\partial {\mathbf{w}}}}\right) ^{T}\right) =\rho
_{i}\theta _{i}\,{\mathbf{\nabla }}\,s_{i}-\rho _{i}\,{\mathbf{\nabla }}%
\,h_{i},
\end{eqnarray*}
and due to the fact that
\begin{equation*}
\mathrm{div\;}{\mathbf{u}}_{i}\ \left( {\frac{\partial \epsilon }{\partial {%
\mathbf{w}}}}\right) ^{T}+\frac{\partial }{\partial \mathbf{x}}\left( {\frac{%
\partial \epsilon }{\partial {\mathbf{w}}}}\right) ^{T}{\mathbf{u}}_{i}=%
\mathrm{div}\left( \left( {\frac{\partial \epsilon }{\partial {\mathbf{w}}}}%
\right) ^{T}\otimes {\mathbf{u}}_{i}\right) ,
\end{equation*}
we get the equations of motion of the two components in the form
\begin{eqnarray}
&&\frac{\partial {\rho _{i}\mathbf{u}}_{i}}{\partial t}\,+\,\mathrm{div}%
(\rho _{i}{\mathbf{u}}_{i}\otimes {\mathbf{u}}_{i})+(-1)^{i}\left( \frac{%
\partial }{\partial t}\left( {\frac{\partial \epsilon }{\partial {\mathbf{w}}%
}}\right) ^{T}+\left( \frac{\partial {\mathbf{u}}_{i}}{\partial {\mathbf{x}}}%
\right) ^{T}\left( {\frac{\partial \epsilon }{\partial {\mathbf{w}}}}\right)
^{T}\right.  \notag \\
&&\left. \ +\ \mathrm{div}\left( \left( {\frac{\partial \epsilon }{\partial {%
\mathbf{w}}}}\right) ^{T}\otimes {\mathbf{u}}_{i}\right) \right) =\rho
_{i}\theta _{i}\,{\mathbf{\nabla }}\,s_{i}-\rho _{i}\,{\mathbf{\nabla }}%
\,h_{i}.  \label{equation of
motions i}
\end{eqnarray}
We consider only the isotropic case where the potential energy $\varepsilon $
of the mixture can be written in terms of the isotropic invariants
\begin{equation*}
\beta _{ij}=\mathbf{\nabla }\rho _{i}\cdot \mathbf{\nabla }\rho _{j},\;\chi
_{ij}=\mathbf{\nabla }\rho _{i}\cdot \mathbf{\nabla }s_{j},\;\gamma _{ij}=%
\mathbf{\nabla }s_{i}\cdot \mathbf{\nabla }s_{j},\ (i,j\ =1,2)\text{\ \ and
\ }\omega =\dfrac{1}{2}{\mathbf{w}}^{2}.
\end{equation*}
\begin{equation*}
\varepsilon =\varepsilon (\rho _{i},s_{i},\beta _{ij},\chi _{ij},\gamma
_{ij},\omega ),
\end{equation*}

Then, the equation of motion of each component of the mixture is
\begin{eqnarray}
&&\frac{\partial {\rho _{i}\mathbf{u}}_{i}}{\partial t}\,+\,\mathrm{div}%
(\rho _{i}{\mathbf{u}}_{i}\otimes {\mathbf{u}}_{i})+  \label{equ of motion i}
\\
&&(-1)^{i}\left( \frac{\partial }{\partial t}(a{\mathbf{w}})+a\left( \frac{%
\partial {\mathbf{u}}_{i}}{\partial {\mathbf{x}}}\right) ^{T}{\mathbf{w}}+%
\mathrm{div}(a\ {\mathbf{w}}\otimes {\mathbf{u}}_{i})\right) =\rho
_{i}\theta _{i}\,{\mathbf{\nabla }}\,s_{i}-\rho _{i}\,{\mathbf{\nabla }}%
\,h_{i},  \notag
\end{eqnarray}
where $a=\dfrac{\partial \varepsilon }{\partial \omega }$.

In this case, $\mathbf{\Psi }_{i}\ $and $\mathbf{\Phi }_{i}$ can be written
\begin{equation*}
\mathbf{\Psi }_{i}=\sum_{j=1}^{2}D_{ij}\mathbf{\nabla }\rho _{j}+E_{ij}%
\mathbf{\nabla }s_{j}\,,\;\;\mathbf{\Phi }_{i}=\sum_{j=1}^{2}C_{ij}\mathbf{%
\nabla }\rho _{j}+D_{ij}\mathbf{\nabla }s_{j}\ ,
\end{equation*}
with $\;C_{ij}=(1+\delta _{ij})\;\epsilon ,_{\beta
_{ij}},\;D_{ij}=\;\epsilon ,_{\chi _{ij}},\ E_{ij}=(1+\delta
_{ij})\;\epsilon ,_{\gamma _{ij}},\ $where $\delta _{ij}\;$is the Kronecker
symbol.

The simplest model is when $C_{ij}=C_{ji},\;D_{ij},\;E_{ij}=E_{ji}$ are
constant. Then,
\begin{equation}
\varepsilon =e(\rho _{i},s_{i},{\mathbf{w)\ +}}\sum_{i,j=1}^{2}\frac{1}{2}%
\,C_{ij}\mathbf{\nabla }\rho _{i}\cdot \mathbf{\nabla }\rho _{j}+D_{ij}%
\mathbf{\nabla }\rho _{i}\cdot \mathbf{\nabla }s_{j}+\frac{1}{2}\,E_{ij}%
\mathbf{\nabla }s_{i}\cdot \mathbf{\nabla }s_{j}  \label{isotropic energy}
\end{equation}
where the associated quadratic form with respect to the vectors $\mathbf{%
\nabla }\rho _{i}$ and $\mathbf{\nabla }s_{i}$ is in the form $%
\sum_{i,j=1}^{2}\frac{1}{2}\,C_{ij}\mathbf{\nabla }\rho _{i}\cdot \mathbf{%
\nabla }\rho _{j}+D_{ij}\mathbf{\nabla }\rho _{i}\cdot \mathbf{\nabla }s_{j}+%
\frac{1}{2}\,E_{ij}\mathbf{\nabla }s_{i}\cdot \mathbf{\nabla }s_{j}$.\ This
quadratic form is assumed positive such as the effect of gradient terms
increases the value of the internal energy with respect to a mixture in a
homogeneous configuration.

\subsection{Equation of total momentum and equation of energy for
conservative motions of thermocapillary mixtures}

We limit first to the conservative case. We notice that the equation of
motion of each component is not in divergence form. Nevertheless, by summing
eqs (\ref{equ of motion i}) with respect to $i$, we obtain the balance
equation for the total momentum in a divergence form. In fact, eqs (\ref{equ
of motion i}) imply
\begin{eqnarray}
&&\left( \sum_{i=1}^{2}\frac{\partial {\rho _{i}\mathbf{u}}_{i}}{\partial t}%
\,+\,\mathrm{div}(\rho _{i}{\mathbf{u}}_{i}\otimes {\mathbf{u}}_{i})\right) -%
\mathrm{div}(a\ {\mathbf{w}}\otimes {\mathbf{w}})=  \notag \\
&&\ a\;\left( \frac{\partial {\mathbf{w}}}{\partial {\mathbf{x}}}\right) ^{T}%
{\mathbf{w}}+\left( \sum_{i=1}^{2}\rho _{i}\theta _{i}\,{\mathbf{\nabla }}%
\,s_{i}-\rho _{i}\,{\mathbf{\nabla }}\,h_{i}\right)  \label{sumi}
\end{eqnarray}
In coordinates, the second member of eq. (\ref{sumi}) is
\begin{equation*}
a\ w_{\nu }w_{\gamma ,\nu }+\sum_{i=1}^{2}\epsilon _{,s_{i}}s_{i,\gamma
}-\left( \epsilon _{,s_{i,\nu }}\right) _{,\nu }s_{i,\gamma }-\rho
_{i}\left( \epsilon _{,\rho _{i}}\right) _{,\gamma }+\rho _{i}\left(
\epsilon _{,\rho _{i,\nu \nu }}\right) _{,\gamma }\ ,
\end{equation*}
where $\nu $ is summed from $1$ to $3$. Noting that
\begin{equation}
\epsilon _{,\gamma }=\sum_{i=1}^{2}\epsilon _{,s_{i}}s_{i,\gamma }+\epsilon
_{,s_{i,\nu }}s_{i,\nu \gamma }+\epsilon _{,\rho _{i}}\rho _{i,\gamma
}+\epsilon _{,\rho _{i,\nu }}\rho _{i,\nu \gamma }+a\ w_{\nu }w_{\gamma ,\nu
}\ ,  \label{potential}
\end{equation}
we obtain
\begin{eqnarray*}
&&a\ w_{\nu }w_{\gamma ,\nu }+\sum_{i=1}^{2}\epsilon _{,s_{i}}s_{i,\gamma
}-\left( \epsilon _{,s_{i,\nu }}\right) _{,\nu }s_{i,\gamma }-\rho
_{i}\left( \epsilon _{,\rho _{i}}\right) _{,\gamma }+\rho _{i}\left(
\epsilon _{,\rho _{i,\nu \nu }}\right) _{,\gamma }=\epsilon _{,\gamma }- \\
&&\sum_{i=1}^{2}\epsilon _{,s_{i,\nu }}s_{i,\nu \gamma }+\epsilon _{,\rho
_{i}}\rho _{i,\gamma }+\epsilon _{,\rho _{i,\nu }}\rho _{i,\nu \gamma
}+\left( \epsilon _{,s_{i,\nu }}\right) _{,\nu }s_{i,\gamma }+\rho
_{i}\left( \epsilon _{,\rho _{i}}\right) _{,\gamma }-\rho _{i}\left(
\epsilon _{,\rho _{i,\nu \nu }}\right) _{,\gamma } \\
&=&\epsilon ,_{\gamma }+\sum_{i=1}^{2}\left( -\rho _{i}\epsilon _{,\rho
_{i}}+\rho _{i}\left( \epsilon _{,\rho _{i,\nu }}\right) _{,\nu }\right)
_{,\gamma }-\left( \Phi _{i\nu }\rho _{i,\gamma }+\Psi _{i\nu }s_{i,\gamma
}\right) _{,\nu }\ ,
\end{eqnarray*}
and consequently the equation of motion for the total momentum is
\begin{equation}
\frac{\partial {\rho \mathbf{u}}}{\partial t}\,+\,\mathrm{div}\left(
\sum_{i=1}^{2}\left( \rho \ {\mathbf{u}}_{i}\otimes {\mathbf{u}}_{i}\right)
-\rho a\ {\mathbf{w}}\otimes {\mathbf{w}}-\mathbf{\sigma }\right) =0
\label{equation mixture}
\end{equation}
where $\rho =\rho _1+\rho _2$ is the total volume mass, ${\rho
\mathbf{u}}= \rho _{1}{\mathbf{u}}_{1} + \rho _{2}{\mathbf{u}}_{2}\
$is the total momentum and$\ \mathbf{\sigma }  = \mathbf{\sigma}_{1}
+ \mathbf{\sigma}_{2}$ is the total stress tensor such that
\begin{equation*}
\sigma _{i\nu \gamma }=(-P_{i}+\rho _{i}\ \mathrm{div\ }\mathbf{\Phi }%
_{i})~\delta _{\nu \gamma }-\Phi _{i\nu }\rho _{i,\gamma }~-\Psi _{i\nu
}s_{i,\gamma },\;\text{with \ }P_{i}=\rho _{i}\epsilon _{,\rho _{i}}-\dfrac{%
\rho _{i}\epsilon }{\rho }.\newline
\end{equation*}
Let us notice that if $\epsilon $ depends also on $\mathbf{(}t\mathbf{,x)}$
corresponding to an external force potential, an additive term appears as
body force in relation (\ref{potential}) and in eq. (\ref{equation mixture})
the body force appears in the second member. This is not the case in eqs (%
\ref{equation of motions i}) which include the body forces coming from $%
\epsilon $ (depending on $\mathbf{(}t\mathbf{,x))}$ in terms $h_{i}$. For
the sake of simplicity, we do not introduce the body force in equation of
the total momentum and equation of total energy.

The equation of energy of the total mixture is obtained in the divergence
form. Let us define
\begin{eqnarray*}
&&\mathbf{M}_{i}=\frac{\partial {\rho _{i}\mathbf{u}}_{i}}{\partial t}\,+\,%
\mathrm{div}(\rho _{i}{\mathbf{u}}_{i}\otimes {\mathbf{u}}_{i})+ \\
&&(-1)^{i}\left( \frac{\partial }{\partial t}(a\ \mathbf{w})+a\ \left( \frac{%
\partial {\mathbf{u}}_{i}}{\partial {\mathbf{x}}}\right) ^{T}\mathbf{w}+%
\mathrm{div}\left( {a}\mathbf{w}\otimes {\mathbf{u}}_{i}\right) \right)
-\rho _{i}\theta _{i}\,{\mathbf{\nabla }}\,s_{i}+\rho _{i}\,{\mathbf{\nabla }%
}\,h_{i}, \\
&&G_{i}={\frac{\partial \rho _{i}}{\partial t}}\ +\ \text{div}(\rho _{i}{%
\mathbf{u}}_{i}), \\
&&S=\sum_{i=1}^{2}\left( {\frac{\partial \rho _{i}s_{i}}{\partial t}}\ +\
\mathrm{div}(\rho _{i}s_{i}{\mathbf{u}}_{i})\right) \theta _{i}, \\
&&E=\frac{\partial }{\partial t}\left( \left( \sum_{i=1}^{2}\frac{1}{2}\rho
_{i}\mathbf{u}_{i}^{2}\right) +\varepsilon -a~\mathbf{w}^{2}\right) + \\
&&\mathrm{div}\left( \left( \sum_{i=1}^{2}\rho _{i}\left( \mathbf{K}%
_{i}\cdot \mathbf{u}_{i}-\frac{1}{2}\mathbf{u}_{i}^{2}-\mathbf{\sigma }%
_{i}\right) \mathbf{u}_{i}\right) +\varepsilon \ \mathbf{u-U}\right),
\end{eqnarray*}
where $\mathbf{U}=\sum_{i=1}^{2}\left( \dfrac{d_{i}\rho _{i}}{dt}\mathbf{%
\Phi }_{i}+\dfrac{d_{i}s_{i}}{dt}\mathbf{\Psi }_{i}\right) \ $corresponds to
the \textit{interstitial working:} in the same way as for\textit{\ Cahn}
\textit{and} \textit{Hilliard fluids}, an additional term that has the
physical dimension of a heat flux must be added to the equation of energy
\cite{Eglit,Dunn,Gouin5,Berdichevsky}.

\QTP{Body Math}
\textsc{Theorem}: \emph{For all motions of a thermocapillary fluid mixture,
the relation}
\begin{equation}
E-S+\sum_{i=1}^{2}\left( \frac{1}{2} \mathbf{u}_{i}^{2}-h_{i}+\theta
_{i}s_{i}\right) \ G_{i}-\mathbf{M}_{i}\cdot \mathbf{u}_{i}\equiv 0
\label{Identity}
\end{equation}
\emph{is identicaly satisfied.}

The proof comes from the following algebraic calculation:
\begin{eqnarray*}
&&\mathbf{M}_{i}^{T}\ \mathbf{u}_{i}-\left( \frac{1}{2}\mathbf{u}%
_{i}^{2}-h_{i}\right) G_{i}+\rho _{i}\theta _{i}\frac{d_{i}s_{i}}{dt}\equiv
\frac{\partial }{\partial t}(\frac{1}{2}\rho _{i}\mathbf{u}_{i}^{2})+\mathrm{%
div}((\frac{1}{2}\rho _{i}\mathbf{u}_{i}^{2}){\mathbf{u}}_{i})+h_{i}G_{i} \\
&&+\rho _{i}\theta _{i}\frac{\partial s_{i}}{\partial t}+\rho _{i}\frac{%
\partial h_{i}}{\partial \mathbf{x}}{\mathbf{u}}_{i}+(-1)^{i}\left( \left(
\frac{\partial a\ \mathbf{w}}{\partial t}\right) ^{T}+a\ \mathbf{w}^{T}\frac{%
\partial {\mathbf{u}}_{i}}{\partial {\mathbf{x}}}+\mathrm{div}\left( {a\
\mathbf{u}}_{i}\mathbf{w}^{T}\right) \right) {\mathbf{u}}_{i}
\end{eqnarray*}
Let us note that
\begin{eqnarray*}
&&\frac{\partial \varepsilon }{\partial t}+\mathrm{div}\left(
\sum_{i=1}^{2}\rho _{i}h_{i}{\mathbf{u}}_{i}\right) \equiv a\ \mathbf{w}^{T}%
\frac{\partial {\mathbf{w}}}{\partial t}+\sum_{i=1}^{2}\varepsilon _{,\rho
_{i}}\frac{\partial \rho _{i}}{\partial t}+\mathrm{div}\left( \frac{\partial
\rho _{i}}{\partial t}\frac{\partial \varepsilon }{\partial \mathbf{\nabla }%
\rho _{i}}\right) \\
&&-\frac{\partial \rho _{i}}{\partial t}\;\mathrm{div}\left( \frac{\partial
\varepsilon }{\partial \mathbf{\nabla }\rho _{i}}\right) +\varepsilon
_{,s_{i}}\frac{\partial s_{i}}{\partial t}+\frac{\partial \varepsilon }{%
\partial \mathbf{\nabla }s_{i}}\mathbf{\nabla }\frac{\partial s_{i}}{%
\partial t}+h_{i}\ \mathrm{div}\left( \rho _{i}{\mathbf{u}}_{i}\right) +\rho
_{i}\frac{\partial h_{i}}{\partial \mathbf{x}}{\mathbf{u}}_{i},
\end{eqnarray*}
yields
\begin{eqnarray*}
&&\sum_{i=1}^{2}h_{i}\ G_{i}\equiv \frac{\partial \varepsilon }{\partial t}%
-a\ \mathbf{w}^{T}\frac{\partial {\mathbf{w}}}{\partial t}+ \\
&&\sum_{i=1}^{2}\ \mathrm{div}\left( \rho _{i}h_{i}{\mathbf{u}}_{i}\right) -%
\mathrm{div}\left( \frac{\partial \rho _{i}}{\partial t}\frac{\partial
\varepsilon }{\partial \mathbf{\nabla }\rho _{i}}\right) -\varepsilon
_{,s_{i}}\frac{\partial s_{i}}{\partial t}-\frac{\partial \varepsilon }{%
\partial \mathbf{\nabla }s_{i}}\mathbf{\nabla }\frac{\partial s_{i}}{%
\partial t}-\rho _{i}\frac{\partial h_{i}}{\partial \mathbf{x}}{\mathbf{u}}%
_{i}.
\end{eqnarray*}
Consequently,
\begin{eqnarray*}
&&\sum_{i=1}^{2}\left( \mathbf{M}_{i}^{T}\ \mathbf{u}_{i}-\left( \frac{1}{2}%
\mathbf{u}_{i}^{2}-h_{i}+\theta _{i}s_{i}\right) G_{i}\right) +S\equiv \\
&&\frac{\partial }{\partial t}\left( \left( \sum_{i=1}^{2}\frac{1}{2}\rho
_{i}\mathbf{u}_{i}^{2}\right) +\varepsilon \right) -a\ \mathbf{w}^{T}\frac{%
\partial {\mathbf{w}}}{\partial t}+\sum_{i=1}^{2}\mathrm{div}\left( \rho
_{i}\left( \frac{1}{2}\mathbf{u}_{i}^{2}+h_{i}\right) \mathbf{u}_{i}\right)
\\
&&-\mathrm{div}\left( \frac{\partial \rho _{i}}{\partial t}\mathbf{\Phi }%
_{i}+\frac{\partial s_{i}}{\partial t}\mathbf{\Psi }_{i}\right)
+(-1)^{i}\left( \left( \frac{\partial a\ {\mathbf{w}}}{\partial t}\right)
^{T}{\mathbf{u}}_{i}+a\ \mathbf{w}^{T}\frac{\partial \mathbf{u}_{i}}{%
\partial \mathbf{x}}\ {\mathbf{u}}_{i}+\mathrm{div}\left( a\ \mathbf{u}_{i}%
\mathbf{w}^{T}\right) \ \mathbf{u}_{i}\right) \equiv \\
&&\frac{\partial }{\partial t}\left( \left( \sum_{i=1}^{2}\frac{1}{2}\rho
_{i}\mathbf{u}_{i}^{2}\right) +\varepsilon -a\ \mathbf{w}^{2}\right)
+\sum_{i=1}^{2}\mathrm{div}\left( \rho _{i}\left( \mathbf{K}_{i}^{T}\,%
\mathbf{u}_{i}-\frac{1}{2}\mathbf{u}_{i}^{2}+h_{i}\right) \mathbf{u}%
_{i}\right) \\
&&-\mathrm{div}\left( \frac{\partial \rho _{i}}{\partial t}\mathbf{\Phi }%
_{i}+\frac{\partial s_{i}}{\partial t}\mathbf{\Psi }_{i}\right)
\end{eqnarray*}
Taking into account of the relations
\begin{equation*}
\frac{\partial \rho _{i}}{\partial t}\mathbf{\Phi }_{i}\equiv \frac{%
d_{i}\rho _{i}}{dt}\mathbf{\Phi }_{i}-\mathbf{\Phi }_{i}\frac{\partial
\mathbf{\rho }_{i}}{\partial \mathbf{x}}{\mathbf{u}}_{i}\ \ \mathrm{and\ \ }%
\frac{\partial s_{i}}{\partial t}\mathbf{\Psi }_{i}\equiv \frac{d_{i}s_{i}}{%
dt}\mathbf{\Psi }_{i}-\mathbf{\Psi }_{i}\frac{\partial \mathbf{s}_{i}}{%
\partial \mathbf{x}}{\mathbf{u}}_{i}
\end{equation*}
and the definition of the total stress tensor,
\begin{equation*}
\mathbf{\sigma }\ \equiv \varepsilon -\sum_{i=1}^{2}\left( \rho
_{i}\,\varepsilon _{,\rho _{i}}-\rho _{i}\,\mathrm{div}\mathbf{\Phi }%
_{i}\right) \mathbf{Id}-\mathbf{\Phi }_{i}\frac{\partial \mathbf{\rho }_{i}}{%
\partial \mathbf{x}}-\mathbf{\Psi }_{i}\frac{\partial \mathbf{s}_{i}}{%
\partial \mathbf{x}},
\end{equation*}
where $\mathbf{Id\ }$is the identity tensor,$\ $we deduce immediately the
algebraic identity (\ref{Identity}).$\ \ \ \ \ \ \ \ \ \ \ \ \ \ \ \ \ \ \ \
\ \ \ \ \ \ \ \ \ \ \ \ \ \ \ \ \ \ \ \ \ \ \ \ \ \ \ \ \ \ \ \ \ \ \ \ \ \
\ \ \ \ \ \ \ \ \ \ \ \ \ \ \ \ \ \ \ \ \ \ \ \ \ \ \ \ \ \ \ \ \ \ \ \
\square $ \newline

\textsc{Corollary}: \emph{All conservative motions of a thermocapillary
mixture satisfy the equation of energy balance}
\begin{eqnarray}
&&\frac{\partial }{\partial t}\left( \left( \sum_{i=1}^{2}\frac{1}{2}\rho
_{i}\mathbf{u}_{i}^{2}\right) +\varepsilon -a~\mathbf{w}^{2}\right) +  \notag
\\
&&\mathrm{div}\left( \left( \sum_{i=1}^{2}\rho _{i}\left( \mathbf{K}%
_{i}\cdot \mathbf{u}_{i}-\frac{1}{2}\mathbf{u}_{i}^{2}-\mathbf{\sigma }%
_{i}\right) \mathbf{u}_{i}\right) +\varepsilon \ \mathbf{u-U}\right) =0.
\label{Total energy}
\end{eqnarray}
This result comes from the simultaneity of relations $G_{i}=0,\
S_{i}=0$ and $\mathbf{M}_{i}=0.$ \newline Let us note that
$\varepsilon -a~\mathbf{w}^{2}$ corresponds to the volume internal
energy $\varpi$ of the mixture.

\subsection{Dissipative motions of thermocapillary mixtures}

The conservative fluid mixture model presented in section 2.1 is relevant to
the so-called \textit{second gradient theory} \cite{Germain}, \cite{Trusdell}%
. In our form of equations of mass conservation for each component ({\ref
{mass0}}), equation of the total momentum of the mixture ({\ref{equation
mixture}), equation of energy (\ref{Total energy}), the diffusion term $%
J=\rho _{1}(\mathbf{u}_{1}-\mathbf{u})$ does not directly appear but is
deduced respectively from the velocities and densities of the components.%
\newline
Our aim is to verify the consistency of our model with the second
law of thermodynamics. The introduction of dissipative forces is
simply done in the framework of \textit{the first gradient theory}
\cite{Germain}: the dissipative forces applied to the continuous
medium are divided into volume forces $\mathbf{f}_{i}^{d}$ and
surface forces associated with the Cauchy stress tensor
$\mathbf{\sigma }_{i}^{d}$. Then, the virtual work of dissipative
forces$\ \delta {T}_{i}\ $applied to the component $i\ $is in the
form
\begin{equation*}
\delta {T}_{i}\ =\ \mathbf{f}_{i}^{d}\cdot \mathbf{\zeta }_{i}\ -\
tr\left( \mathbf{\sigma }_{i}^{d}\ {\frac{\partial \mathbf{\zeta }_{i}}{%
\partial \mathbf{x}}}\right),
\end{equation*}
where $\delta {T}_{i}\ $is a differential form. For such dissipative
motions, no production of masses due to chemical reactions appears.\newline
For the same virtual displacement of two components, $\ \mathbf{\zeta }\ =\
\mathbf{\zeta }_{i},\ \ (i=1,2)\ $, the total virtual work of dissipative
forces is
\begin{equation*}
\delta {T}\ =\ \sum_{i=1}^{2}\ \mathbf{f}_{i}^{d}\cdot \mathbf{\zeta }\ -\
tr \left( \mathbf{\sigma }_{i}^{d}\ {\frac{\partial \mathbf{\zeta }}{%
\partial \mathbf{x}}}\right).
\end{equation*}
When $\mathbf{\zeta \ }$is a translation, the work $\delta {T}$ is equal to
zero and consequently,
\begin{equation}
\sum_{i=1}^{2}\ \ \mathbf{f}_{i}^{d}=\ 0\ \ \ \mathrm{or}\ \ \ \mathbf{f}%
_{2}^{d}=-\mathbf{f}_{1}^{d}\equiv \mathbf{f}^{d}  \label{diffusion}
\end{equation}
We specify later the behavior of forces $\mathbf{f}_{i}^{d}\ $ (they will be
associated with the diffusion term) and stress tensors $\mathbf{\sigma }%
_{i}^{d}$.\ Taking into account of the dissipative effects, the
equations of motion for each component become
\begin{eqnarray}
\frac{\partial {\rho _{i}\mathbf{u}}_{i}}{\partial t}\,+\,\mathrm{div}(\rho
_{i}{\mathbf{u}}_{i}\otimes {\mathbf{u}}_{i}) &+&(-1)^{i}\left( \frac{%
\partial }{\partial t}(a{\mathbf{w}})+a\ \left( \frac{\partial {\mathbf{u}}%
_{i}}{\partial {\mathbf{x}}}\right) ^{T}{\mathbf{w}}+\mathrm{div}(a\ {%
\mathbf{w}}\otimes {\mathbf{u}}_{i})\right)   \notag \\
&=&\rho _{i}\theta _{i}\,{\mathbf{\nabla }}\,s_{i}-\rho _{i}\,{\mathbf{%
\nabla }}\,h_{i}+\,\mathrm{div\ }\mathbf{\sigma }_{i}^{d}+\mathbf{f}_{i}^{d}
\end{eqnarray}
Taking into account of relation (\ref{diffusion}), the equation of the total
momentum writes
\begin{equation}
\frac{\partial {\rho \mathbf{u}}}{\partial t}\,+\,\mathrm{div}\left(
\sum_{i=1}^{2}\left( \rho \ {\mathbf{u}}_{i}\otimes {\mathbf{u}}_{i}\right)
-\rho a\ {\mathbf{w}}\otimes {\mathbf{w}}-\mathbf{\sigma -\sigma }^{d}\
\right) =0  \label{dissipative momentum}
\end{equation}
with $\mathbf{\sigma }^{d}=\mathbf{\sigma }_{1}^{d}+\mathbf{\sigma }_{2}^{d}.
$\newline
The introduction of the heat flux vector $\mathbf{q}$ and the heat supply $r$
comes from classical methods in thermodynamics \cite{ET,Muller,Groot}. If we
write,
\begin{eqnarray*}
&&\mathbf{M}_{i}^{d}=\mathbf{M}_{i}-\mathrm{div\ }\mathbf{\sigma }_{i}^{d}-\
\mathbf{f}_{i}^{d},\ \  \\
&&S^{d}=S-r+\mathrm{div\ }\mathbf{q}+\sum_{\alpha =1}^{2}\ \mathbf{f}%
_{i}^{d}\ .\mathbf{u}_{i}-tr\ (\mathbf{\sigma }_{i}^{d}\ \mathbf{\Delta }%
_{i})\mathbf{,}\ \ \  \\
&&E^{d}=E-r+\mathrm{div\ }\mathbf{q}-\sum_{\alpha =1}^{2}\mathrm{div\ }%
\mathbf{\sigma }_{i}^{d}{\mathbf{u}}_{i},
\end{eqnarray*}
with $\mathbf{\Delta }_{i}=\dfrac{1}{2}\left( \dfrac{\partial \mathbf{u}_{i}%
}{\partial \mathbf{x}}+\left( \dfrac{\partial \mathbf{u}_{i}}{\partial
\mathbf{x}}\right) ^{T}\right) $ represents the velocity deformation tensor
of each component, relation (\ref{Identity}) writes as
\begin{equation}
E^{d}-S^{d}-\sum_{i=1}^{2}\mathbf{M}_{i}^{d}\cdot \mathbf{u}_{i}-\left(
\mathbf{K}_{i}\cdot \mathbf{u}_{i}-R_{i}-\theta _{i}s_{i}\right) \
G_{i}\equiv 0  \label{Gibbs}
\end{equation}
which can be considered as the dynamic form of the \emph{Gibbs identity}.%
\newline
For the components of the mixture, equations of momenta and equations of
masses are in the form
\begin{equation}
\mathbf{M}_{i}^{d}=0,\ G_{i}=0\,.  \label{Bilan}
\end{equation}
The \emph{Gibbs identity} (\ref{Gibbs}) and eqs (\ref{Bilan}) imply $%
S^{d}=E^{d}$.$\ $If we assume that\ $S^{d}=0$, i.e.,
\begin{equation}
\sum_{i=1}^{2}\ \left( {\frac{\partial \rho _{i}s_{i}}{\partial t}}\ +\
\mathrm{div}(\rho _{i}s_{i}{\mathbf{u}}_{i})\right) \theta _{i}+\mathbf{f}%
_{i}^{d}\ .\mathbf{u}_{i}-tr\ (\mathbf{\sigma }_{i}^{d}\ \mathbf{\Delta }%
_{i})-r+\mathrm{div\ }\mathbf{q=0,}  \label{entropy diss}
\end{equation}
it is equivalent to write $E^{d}=0$, i.e.,
\begin{eqnarray}
&&\frac{\partial }{\partial t}\left( \left( \sum_{i=1}^{2}\frac{1}{2}\rho
_{i}\mathbf{u}_{i}^{2}\right) +\varepsilon -a~\mathbf{w}^{2}\right) +
\label{dissipative energy} \\
&&\mathrm{div}\left( \left( \sum_{i=1}^{2}\rho _{i}\left( \mathbf{K}%
_{i}\cdot \mathbf{u}_{i}-\frac{1}{2}\mathbf{u}_{i}^{2}-\mathbf{\sigma }_{i}-%
\mathbf{\sigma }_{i}^{d}\right) \mathbf{u}_{i}\right) +\varepsilon \ \mathbf{%
u-U+q}\right) -r=0.  \notag
\end{eqnarray}
Eq. (\ref{entropy diss}) is the \textit{equation of entropy} and Eq. (\ref
{dissipative energy}) is the \textit{equation of energy.} }

In the conservative case, the system is closed with two different
temperatures $\theta _{i}\ (i=1,2)$. In the dissipative case we need
additional arguments to obtain equations for each entropy $s_{i}$
which could remplace equations (\ref{entropy i}). A possibility is
to consider the case when the exchanges of momentum and energy
between the two components are rapid enough to have a common
temperature (this is not the case of heterogeneous mixtures where
each phase may have different pressures and temperatures
\cite{lhuillier}). Then, in dissipative case, if we know all the
dissipative functions, the governing system is closed. Note also
that it could be possible to consider a common temperature and
entropy both for conservative and dissipative case. This case is
connected with a conservative equation for the common entropy $s$
(see Appendix). Another possibility is to assume that the entropy is
transported along the $i$th component (say, for example $i=1$) which
was used for quantum fluids by Landau \cite{Landau}
\begin{equation*}
{\frac{\partial \rho _{1}s}{\partial t}}\ +\ div\ (\rho _{1}s\ \mathbf{u}%
_{1})\ =\ 0.
\end{equation*}
In this case, the independent functions are $\rho _{1},s,\mathbf{u}_{1},\rho
_{2},\mathbf{u}_{2}$, where $\rho _{i}$ are submitted to the constraints (%
\ref{mass i}) and the case of Helium superfluid is a special case of our
study corresponding to $s_{2}=0$ and $s=s_{1}$. Using of this hypothesis is
nevertheless doubtful for classical fluids.

Hence, we may suppose a common temperature only for dissipative case. This
means that
\begin{equation}
\theta =\theta _{1}=\theta _{2}  \label{Temperature}
\end{equation}
This hypothesis closes the system (\ref{Bilan}-\ref{entropy diss}).

\textit{Now we focus on the governing equations for each components of the
mixture}:
\begin{equation*}
\mathbf{M}_{i}^{d}=0 \ \ \ (i=1,2).
\end{equation*}
For slow motions, we rewrite these equations in the following form :
\begin{equation*}
\mathbf{M}_{i}^{d} \simeq \rho _{i}\,{\mathbf{\nabla }} \,h_{i} -\rho _{i}\,
\theta \,{\mathbf{\nabla }}\,s_{i}-\mathrm{div\ }\mathbf{\sigma }_{i}^{d}-\
\mathbf{f}_{i}^{d}= 0.
\end{equation*}
If we consider the case when the motion of each component is regular enough,
an approximative case is the case of \textit{solid displacements}\, for the
motion of each component; then, $\mathrm{div \ }\mathbf{\sigma }%
_{i}^{d}\simeq 0$ and consequently,
\begin{equation*}
\mathbf{M}_{i}^{d} \simeq \rho _{i}\,{\mathbf{\nabla }} \,\mu_{i} -\ \mathbf{%
f}_{i}^{d}= 0,
\end{equation*}
where $\mu_i = h_i-\theta s_i$ is the chemical potential of the component $i$
of the mixture at the temperature $\theta$.\newline
Considering the difference $\mathbf{M}_{2}^{d}-\mathbf{M}_{1}^{d}$ and using
relation (\ref{diffusion}), we obtain,
\begin{equation}
\frac{\mathbf{f}_{2}^{d}}{\rho_2}-\frac{\mathbf{f}_{1}^{d}}{\rho_1}\equiv
\frac{\rho\, \mathbf{f}^{d}}{\rho_1\rho_2} = {\mathbf{\nabla }} \,\mu,
\label{gradchem}
\end{equation}
where $\mu = \mu_2 - \mu_1$ is the chemical potential of the mixture \cite
{Bowen,Nigmatulin}. Let us introduce the diffusion flux $\mathbf{J}$,
\begin{equation}
\mathbf{J} \equiv \rho_1 (\mathbf{u}_1-\mathbf{u}) = \frac{\rho_1\rho_2}{\rho%
}\mathbf{w}.  \label{diflux}
\end{equation}
Equation (\ref{gradchem}) implies,
\begin{equation}
\mathbf{f}^{d} \cdot \mathbf{w} = {\mathbf{\nabla }} \,\mu \cdot \mathbf{J.}
\label{difentrop}
\end{equation}

The term ${\mathbf{\nabla }} \,\mu \cdot \mathbf{J}$ corresponds to the
entropy production due to the diffusion process. Expression (\ref{difentrop}%
) is the connection between the mechanical drag force between components of
the mixture and the thermodynamical process of diffusion; consequently, Eq. (%
\ref{entropy diss}) yields
\begin{equation*}
\sum_{i=1}^{2}\ \left( {\frac{\partial \rho _{i}s_{i}}{\partial t}}\ +\
\mathrm{div}(\rho _{i}s_{i}{\mathbf{u}}_{i})\right) \theta _{i}-tr\ (\mathbf{%
\sigma }_{i}^{d}\ \mathbf{\Delta }_{i})-r+\mathrm{div\ }\mathbf{q}-{\mathbf{%
\nabla }} \,\mu \cdot \mathbf{J}=0.
\end{equation*}
We have obtained the equations of balance of masses, equation of
total momentum, Eq. (\ref{gradchem}) between the components and
equation of total energy by using an energetic method. They are the
extension of classical mixture equations to equations of mixtures of
fluids involving density gradients.

\subsection{Second law of thermodynamics}

Relation (\ref{entropy diss}) may be rewritten in the form
\begin{equation*}
\ \left( \sum_{i=1}^{2}{\frac{\partial \rho _{i}s_{i}}{\partial t}}\ +\
\mathrm{div}(\rho _{i}s_{i}{\mathbf{u}}_{i})\right) +\mathrm{div\ }\frac{%
\mathbf{q}}{\theta }-\frac{r}{\theta }=\frac{1}{\theta }\left(
\sum_{i=1}^{2}tr\ (\mathbf{\sigma }_{i}^{d}\ \mathbf{\Delta
}_{i})\right)+\mathbf{f}
^{d}\cdot \mathbf{w}-\mathbf{\nabla }\theta \cdot \frac{%
\mathbf{q}}{\theta ^{2}}
\end{equation*}
This last equation yields the entropy production due to the diffusion,
viscosity and heat flux processes. Due to $\theta >0$, if we assume
\begin{equation}
\left( \sum_{i=1}^{2}tr\ (\mathbf{\sigma }_{i}^{d}\ \mathbf{\Delta
}_{i})\right)+
\mathbf{f}^{d}\cdot \mathbf{w} -\nabla \theta \cdot \frac{%
\mathbf{q}}{\theta }\geq 0,  \label{inequality}
\end{equation}
we get the Clausius-Duhem inequality in the form
\begin{equation*}
\ \left( \sum_{i=1}^{2}{\frac{\partial \rho _{i}s_{i}}{\partial t}}\ +\
\mathrm{div}(\rho _{i}s_{i}{\mathbf{u}}_{i})\right) +\mathrm{div\ }\frac{%
\mathbf{q}}{\theta }-\frac{r}{\theta }\geq 0.
\end{equation*}
Let us notice that (\ref{inequality}) is satisfied if $tr\ (\mathbf{\sigma }%
_{i}^{d}\ \mathbf{\Delta }_{i})\geq 0,\ \mathbf{f}^{d}\cdot \mathbf{w}%
\geq 0$ and $\mathbf{\nabla }\theta \cdot\, \mathbf{q}%
\leq 0$. The following constitutive laws are classically adopted for $%
\mathbf{\sigma }_{i}^{d},\ \mathbf{f}^{d}$ and $\mathbf{q:}$

The stress tensor $\mathbf{\sigma }_{i}^{d}\ $is a symmetric isotropic
tensor function of $\mathbf{\Delta }_{i}$ such that $\sum_{i=1}^{2}tr\ (%
\mathbf{\sigma }_{i}^{d}\ \mathbf{\Delta }_{i})\geqslant 0\ $(a special case
of such a function is $\mathbf{\sigma }_{i}^{d}=\lambda _{i}(tr\ \mathbf{%
\Delta }_{i})\mathbf{Id}+2\mu _{i}\mathbf{\Delta }_{i}$ with $\mu
_{i}\geqslant 0$ and $3\lambda _{i}+2\mu _{i}\geqslant 0)$.

The heat flux vector satisfies the Fourier law
\begin{equation*}
\mathbf{q=-}\chi \mathbf{\ \nabla }\theta \text{ \ with \ }\chi \geqslant 0.
\end{equation*}
The linear approximation known in the literature as the \textit{Stokes drag
formula }is adopted \cite{Landau,Bowen}
\begin{equation*}
\mathbf{f}^{d}=k\, \mathbf{w,\ }k \geqslant 0,\ {\rm with} \ \
\mathbf{f}_{2}^{d}=- \mathbf{f}_{1}^{d}\equiv \mathbf{f}^{d}\ \ {\rm
and} \  \ \mathbf{w}\ = \mathbf{u_1}- \mathbf{u_2}.
\end{equation*}

Let us note that relations (\ref{gradchem}) and (\ref{diflux}) together with
the drag formula yield the property of the diffusion flux
\begin{equation}
\mathbf{J} =  \frac{1}{k} \, {\mathbf{\nabla }} \,\mu, \label{Fick}
\end{equation}
which is the general form of the Fick law \cite{Groot,Landau}. So, the Fick
law is not directly a linear phenomenological law as the Fourier law but a
direct consequence of equations of motion and the Stokes drag force
hypothesis which was previously noticed by Bowen for a different model \cite
{Bowen}.

\section{Acceleration Waves}

Now, we will consider conservative motions of thermocapillary mixtures only.%
\newline
As it is well known, wave phenomena - in particular discontinuity waves
(waves across the front of which some derivatives of the field variables
have jumps) - are typical of models that are described through hyperbolic
differential systems. A classical example of discontinuity waves in
continuum mechanics are the so-called \textit{acceleration waves} in which,
among the other variables, the acceleration jumps across the front while the
velocity is continuous \cite{whitham}.

Dissipative systems and in particular models for diffusive processes have
usual differential system with a parabolic structure and discontinuity waves
are not admissible. A typical example of non admissibility is the one of
Navier-Stokes-Fourier fluids. In this case a possible approach to obtain
hyperbolic system is the method of the Extended Thermodynamic theory \cite
{ET}, valid also for rarefied gases.

Nevertheless in parabolic systems some discontinuity waves may propagate for
particular initial data. The aim of this paper is to prove that for the
present model of thermocapillarity fluid binary mixtures there exists the
possibility of propagation of two tangential acceleration waves, provided
that the internal energy is at least a function of the entropy gradient of
one component. For this aim, we first briefly recall some very well known
questions about discontinuity waves.\newline
A wave is a discontinuity wave if the wave front with Cartesian equation $%
\phi (t,\mathbf{x})=0$ separates the space in two subspaces in which there
exists regular solutions of the differential system but across the normal
direction of the front some derivatives of the field suffers a jump \cite
{Ruggeri}.\newline
As usual, we indicate the jump with a square bracket,
\begin{equation*}
\lbrack \ \ ]\ =(\ \ )_{\phi =0^{-}}-(\ \ )_{\phi =0^{+}}
\end{equation*}
and we introduce the map between $(t,\mathbf{x})$ and $(\phi ,\mathbf{\xi })$
, where $\mathbf{\xi \equiv \xi }$ $(t,\mathbf{x})$ represents the
tangential manifold of the wave surface in time-space. Therefore the
assumptions for the discontinuity waves are expressed for a generic function
$f$ in the form,
\begin{equation}
\left\{
\begin{array}{c}
\lbrack f]=0;\ \ \ \ \left[ \dfrac{\partial ^{k}f}{\partial \xi
_{\gamma_{1}}\cdots \partial \xi_{\gamma_{k}}}\right] =0\ \ \forall k; \\
\\
\text{where there exists }p\geqq 1\text{ \ such that} \\
\\
\left[ \dfrac{\partial ^{j}f}{\partial \phi ^{j}}\right] =0\ \ \ \text{for }%
\ \ 1\leq j\leq p-1;\ \ \  \\
\\
\ \ \delta ^{k}f\ \equiv \ \left[ \dfrac{\partial ^{k}f}{\partial \phi ^{k}}%
\right] \neq 0\ \ \ \text{for\ \ \ }k\geq p.
\end{array}
\right.  \label{jump1}
\end{equation}
Taking into account (\ref{jump1})\ and the Hadamard lemma \cite{hadamard},
we have
\begin{equation*}
\left[ \frac{\partial ^{p}f}{\partial x_{\gamma_{1}}\cdots \partial x_{\gamma_{p}}}%
\right] =\delta ^{p}f\ n_{i_{1}}\cdots n_{i_{p}};\ \ \ \ \ \ \ \ \ \ \left[
\frac{\partial ^{p}f}{\partial t^{p}}\right] =(-\lambda )^{p}\delta ^{p}f
\end{equation*}
where $\lambda $ and here $\mathbf{n\equiv (}n_{i})$ are respectively the
normal velocity and the unit normal vector to the wave front.

The advantages of the previous symbols are that there exists a chain rule
between the field derivative in the differential systems and the
corresponding jump relation,
\begin{equation}
\partial _{t}\rightarrow -\lambda \ \delta ;\ \ \ \ \ \ \ \ \ \partial _{\gamma}\
\rightarrow n_{\gamma}\ \delta \ .  \label{jump3}
\end{equation}

We apply now this procedure to our differential systems assuming that across
the wave front $\rho _{i}$,\ $s_{i}$ and its first derivatives are
continuous and there are jumps for the second derivative ($p=2$) while the
velocity is continuous and suffer a jump in the first derivative ($p=1$)
\begin{eqnarray}
\lbrack \rho _{i}] &=&[s_{i}]=[\delta \rho _{i}]=[\delta s_{i}]=0;\ \ \ \
[\delta ^{2}\rho _{i}]\neq 0;\ \ \ [\delta ^{2}s_{i}]\neq 0;\ \   \notag \\
&&\   \label{jump4} \\
\lbrack \mathbf{u}_{i}] &=&0;\ \ \ \ [\delta \mathbf{u}_{i}]\neq 0;\ \ \ \ \
\ \ i=1,2.  \notag
\end{eqnarray}

\subsection{Jump conditions:}

From the balance of mass of the two components (\ref{mass i}),
taking into account of eqs (\ref{jump4}) and the chain rule
(\ref{jump3}), we obtain that the normal components of the first
derivative of the velocities are continuous,
\begin{equation}
\ [\delta u_{in}]=0,\ \ \ \   \label{jump5}
\end{equation}
where $\ u_{in}\ =\mathbf{n\cdot u}_{i}.$ \newline
If we differentiate the entropy balance law of each component (\ref{entropy
i}), with respect $\mathbf{x}$, we obtain for the discontinuities
\begin{equation}
-v_{i}\delta ^{2}s_{i}+\mathbf{\nabla }s_{i}\cdot \delta \mathbf{u}_{i}=0,
\label{jump6}
\end{equation}
where
\begin{equation*}
v_{i}=\lambda -u_{in}
\end{equation*}
are the relative velocities of the wave front with respect to the fluid
components. From the balance of momentum we obtain
\begin{eqnarray}
&&-\rho _{i}v_{i}\delta \mathbf{u}_{i}+(-1)^{i}a\left\{ v_{i}\delta (\mathbf{%
u}_{2}-\mathbf{u}_{1})+\left( \delta \mathbf{u}_{i}\cdot (\mathbf{u}_{1}-%
\mathbf{u}_{2})\right) \;\mathbf{n}\right\}  \notag \\
&&\;\;\;\;\;\;\;\;\;\;\;\;\;\;\;\ \ \ \ \ \ \ \ \ -\rho _{i}\left\{ \mathbf{%
\nabla }s_{i}[\theta _{i}]-[\mathbf{\nabla }h_{i}]\right\} =0.  \label{jump8}
\end{eqnarray}
From eq. (\ref{temperature&enthalpie}) we obtain
\begin{equation*}
B_{i}\equiv -\rho _{i}[\theta _{i}]=\sum_{j=1}^{2}\left( D_{ji}\ \delta
^{2}\rho _{j}+E_{ji}\ \delta ^{2}s_{j}\right)
\end{equation*}
\begin{equation}
\mathbf{A}_{i}\equiv -[\mathbf{\nabla }h_{i}]=\mathbf{n\ }%
\sum_{j=1}^{2}\left( C_{ij}\ \delta ^{3}\rho _{j}+D_{ij}\ \delta
^{3}s_{j}\right) .  \label{Ai}
\end{equation}
Then relation (\ref{jump8}) becomes
\begin{equation*}
-\rho _{i}v_{i}\delta \mathbf{u}_{i}+(-1)^{i}a\left\{ v_{i}\delta (\mathbf{u}%
_{2}-\mathbf{u}_{1})+\left( \delta \mathbf{u}_{i}\cdot (\mathbf{u}_{1}-%
\mathbf{u}_{2})\right) \;\mathbf{n}\right\} +B_{i}\mathbf{\nabla }s_{i}\
-\rho _{i}\mathbf{A}_{i}=0.
\end{equation*}
If we multiply by $\mathbf{n}$ and we take into account of relation (\ref
{jump5}), we obtain
\begin{equation}
\rho _{i}\mathbf{A}_{i}\cdot \mathbf{n}=(-1)^{i}a\ \delta \mathbf{u}%
_{i}\cdot (\mathbf{u}_{1}-\mathbf{u}_{2})+B_{i}\mathbf{\nabla }s_{i}\cdot
\mathbf{n},  \label{third}
\end{equation}
then, we get the final jump conditions from the momentum equations
\begin{equation}
\rho _{i}v_{i}\delta \mathbf{u}_{i}+(-1)^{i}av_{i}\delta (\mathbf{u}_{1}-%
\mathbf{u}_{2})-B_{i}\mathbf{\nabla }_{t}s_{i}=0,  \label{finamom}
\end{equation}
where
\begin{equation}
\mathbf{\nabla }_{t}s_{i}=\mathbf{\nabla }s_{i}-(\mathbf{\nabla }s_{i}\cdot
\mathbf{n})\mathbf{n}  \label{tangential}
\end{equation}
denotes the tangential component of the gradient of entropy of each
component.

Therefore we obtain the algebraic system of \ $8$ equations (\ref{jump5}), (%
\ref{jump6}) and (\ref{finamom}) for the $10$ scalar unknowns $\delta
^{2}\rho _{i},\ $ $\delta ^{2}s_{i}\ $and $\delta \mathbf{u}_{i}\ (i=1,2).\ $%
Consequently, we needs two more conditions that are obtained by
compatibility conditions coming from boundary conditions associated with the
equation of motion for each component. In fact, we notice in Appendix that
if $\rho _{i},s_{i},\mathbf{\nabla }\rho _{i},\mathbf{\nabla }s_{i}$ are
continuous through a surface of weak discontinuities, then \ $\mathrm{div\ }%
\mathbf{\Phi }_{i}\ $must be also continuous through the surface
\begin{equation*}
\left[ \text{div }\mathbf{\Phi }_{i}\right] =0.
\end{equation*}
We notice additively that these two conditions are compatible with the
Rankine-Hugoniot conditions associated to the total momentum (\ref{equation
mixture}) and the total energy balance law (\ref{Total energy}): In fact,
eq. (\ref{equation mixture}) yields
\begin{equation*}
\lbrack \mathbf{\sigma }_{1}+\mathbf{\sigma }_{2}]=0,
\end{equation*}
while from eq. (\ref{Total energy}), we get
\begin{equation*}
\lbrack \mathbf{\sigma }_{1}u_{1n}+\mathbf{\sigma }_{2}u_{2n}]=0.
\end{equation*}
Then, we obtain the two supplementary equations
\begin{equation}
\lbrack \text{div }\mathbf{\Phi }_{i}]=\ \sum_{j=1}^{2}\left( C_{ij}\ \delta
^{2}\rho _{j}+D_{ij}\ \delta ^{2}s_{j}\right) =0\ \   \label{sigmai}
\end{equation}
and the system for the discontinuities becomes an homogeneous closed system
of $10$ equations for $10$ scalar unknowns $\delta ^{2}\rho _{i},\ $ $\delta
^{2}s_{i}\ $and $\delta \mathbf{u}_{i}\ (i=1,2)$ in the form,
\begin{equation}
\left\{
\begin{array}{l}
\lbrack \delta u_{in}]=0 \\
\\
\rho _{i}v_{i}\delta \mathbf{u}_{i}+(-1)^{i}av_{i}\delta (\mathbf{u}_{1}-%
\mathbf{u}_{2})-\sum_{j=1}^{2}\left( D_{ji}\ \delta ^{2}\rho _{j}+E_{ji}\
\delta ^{2}s_{j}\right) \mathbf{\nabla }_{t}s_{i}=0 \\
\\
-v_{i}\delta ^{2}s_{i}+\mathbf{\nabla }s_{i}\cdot \delta \mathbf{u}_{i}=0 \\
\\
\sum_{j=1}^{2}\left( C_{ij}\ \delta ^{2}\rho _{j}+D_{ij}\ \delta
^{2}s_{j}\right) =0
\end{array}
\right.  \label{sistema}
\end{equation}
We observe that the conditions (\ref{Ai}), (\ref{third}) are constraints for
the jump of the third derivatives of densisties and entropies $\delta
^{3}\rho _{j},\ \delta ^{3}s_{j}$.

Now we consider the weak discontinuities near the equilibrium of the fluid
mixture; then
\begin{equation*}
\mathbf{u}_{i}=0.
\end{equation*}
Consequently $v=\lambda $ is the velocity of the acceleration wave. Due to
the fact that thermocapillary mixtures can be considered as a mathematical
model for interfacial layers between two mixture bulks \cite{gouin3}, the
gradients of tensorial quantities $\rho _{i}$ and $s_{i}\ $are orthogonal to
the interfacial layers and consequently are collinear,
\begin{equation*}
\mathbf{\nabla }s_{1}=b\ \mathbf{\nabla }s_{2}.
\end{equation*}
Then the second and the third equations of system (\ref{sistema}) allow to
eliminate $\mathbf{u}_{i}$ and to get
\begin{equation*}
c_{s}^{2}(\left( a-\rho _{1}\right) \ \delta ^{2}s_{1}-a\ b\ \delta
^{2}s_{2})+\sum_{j=1}^{2}\left( D_{j1}\ \delta ^{2}\rho _{j}+E_{j1}\ \delta
^{2}s_{j}\right) (\mathbf{\nabla }_{t}s_{1})^{2}=0
\end{equation*}
and
\begin{equation*}
c_{s}^{2}\left( -a\ b\ \delta ^{2}s_{1}+b^{2}(a\ -\rho _{2})\ \delta
^{2}s_{2}\right) +\sum_{j=1}^{2}\left( D_{j2}\ \delta ^{2}\rho _{j}+E_{j2}\
\delta ^{2}s_{j}\right) \left( \mathbf{\nabla }_{t}s_{1}\right) ^{2}=0
\end{equation*}
where
\begin{equation*}
c_{s}^{2}=\dfrac{v^{2}}{(\mathbf{\nabla }_{t}s_{1})^{2}}.
\end{equation*}
Therefore we obtain a system of compatibility between the variables $\delta
^{2}\rho _{j}$ and $\delta ^{2}s_{j}$ in the form
\begin{equation}
\left\{
\begin{array}{l}
C_{11\ }\delta ^{2}\rho _{1}+C_{12\ }\delta ^{2}\rho _{2}+D_{11}\ \delta
^{2}s_{1}+D_{12}\ \delta ^{2}s_{2}=0 \\
\\
C_{12\ }\delta ^{2}\rho _{1}+C_{22\ }\delta ^{2}\rho _{2}+D_{21}\ \delta
^{2}s_{1}+D_{22}\ \delta ^{2}s_{2}=0 \\
\label{sistema 2} \\
D_{11\ }\delta ^{2}\rho _{1}+D_{21\ }\delta ^{2}\rho _{2}+\left(
E_{11}+\left( a-\rho _{1}\right) c_{s}^{2}\right) \ \delta ^{2}s_{1} \\
\ \ \ \ \ \ \ \ \ \ \ \ \ \ \ \ \ \ \ \ \ \ \ \ \ \ \ +\left( E_{12}-a\ b\
c_{s}^{2}\right) \ \delta ^{2}s_{2}=0 \\
\\
D_{12\ }\delta ^{2}\rho _{1}+D_{22\ }\delta ^{2}\rho _{2}+\left( E_{12}-a\
b\ c_{s}^{2}\right) \ \delta ^{2}s_{1} \\
\ \ \ \ \ \ \ \ \ \ \ \ \ \ \ \ \ \ +\left( E_{22}+\left( a-\rho _{2}\right)
b^{2}c_{s}^{2}\right) \ \delta ^{2}s_{2}=0
\end{array}
\right.
\end{equation}
Let us denote
\begin{eqnarray*}
\mathbf{C} &=&\mathbf{C}^{T}=\left|
\begin{array}{cc}
C_{11} & C_{12} \\
C_{12} & C_{22}
\end{array}
\right| ,\ \mathbf{D}=\left|
\begin{array}{cc}
D_{11} & D_{12} \\
D_{21} & D_{22}
\end{array}
\right| , \\
&& \\
\ \mathbf{E} &=&\mathbf{E}^{T}=\left|
\begin{array}{cc}
E_{11} & E_{12} \\
E_{12} & E_{22}
\end{array}
\right| ,\ \mathbf{B}=\mathbf{B}^{T}=\left|
\begin{array}{cc}
\rho _{1}-a & ab \\
ab & (\rho _{2}-a)b^{2}
\end{array}
\right|
\end{eqnarray*}
\begin{equation*}
\delta ^{2}\mathbf{\rho }=\left|
\begin{array}{c}
\delta ^{2}\rho _{1} \\
\delta ^{2}\rho _{2}
\end{array}
\right| ,\ \delta ^{2}\mathbf{s}=\left|
\begin{array}{c}
\delta ^{2}s_{1} \\
\delta ^{2}s_{2}
\end{array}
\right|
\end{equation*}
and system (\ref{sistema 2})\ is writting
\begin{equation*}
\left\{
\begin{array}{c}
\mathbf{C}\ \delta ^{2}\mathbf{\rho }+\mathbf{D}\ \delta ^{2}\mathbf{s}=0 \\
\mathbf{D}^{T}\ \delta ^{2}\mathbf{\rho }+(\mathbf{E}-c_{s}^{2}\mathbf{B})\
\delta ^{2}\mathbf{s}=0
\end{array}
\right. ,
\end{equation*}
which implies
\begin{equation}
\left( \mathbf{A}-c_{s}^{2}\mathbf{B}\right) \delta ^{2}\mathbf{s}=0.
\label{final}
\end{equation}
where
\begin{equation*}
\mathbf{A}=E-\mathbf{D}^{T}\ \mathbf{C}^{-1}\mathbf{D}\ .
\end{equation*}
From eq.(\ref{final}) it is simple to verify the property,

\textsc{Theorem}: \emph{\ }\textit{If }$\mathbf{EC-D}^{2}$ is \textit{%
positive definite and if we consider small diffusion, i.e.} $a<\rho ^{\ast }$
\textit{with}
\begin{equation*}
\rho ^{\ast }=\frac{\rho _{1}\rho _{2}}{\rho _{1}+\rho _{2}}
\end{equation*}
\textit{then,} $\mathbf{A}$ \textit{and} $\mathbf{B}$ \textit{are both
symmetric and definite positive, all the eigenvalues} $c_{s}^{2}$ \textit{of}
eq. (\ref{final}) \textit{are positive and two discontinuity waves exist.}

\subsection{Exceptional waves:}

As in the hyperbolic case, a wave is \textit{exceptional or linearly
degenerate} (see e.g. \cite{Ruggeri}) if
\begin{equation}
\delta \lambda \equiv 0.  \label{exceptional}
\end{equation}

In this case the wave behavior is similar to the behavior in linear case and
we do not get any distortion of the wave or shock formation.

It is simple matter to prove that both the waves fulfill the exceptionality
condition. In fact, taking into account of relations (\ref{jump3}),(\ref
{tangential}), we have
\begin{equation*}
\delta \mathbf{\nabla }_{t}s\equiv 0
\end{equation*}
and from (\ref{final}) we obtain that $\lambda $ is function of the modulo
of the tangential gradient of entropy
\begin{equation*}
\lambda \equiv \lambda (\left| \mathbf{\nabla }_{t}s\right| )
\end{equation*}
and then (\ref{exceptional}) holds.\newline

\section{Results and discussion:}

In this paper we prove that the model of thermocapillary fluid
mixtures with dissipation yields a system of equations of motions
compatible with the second law of thermodynamics at least in simple
dissipative cases. The equation of motion and the equation of energy
of the barycentric motion of the mixture are in a divergence form in
conservative cases.\newline Consequently, Hamilton's principle
applied to fluid dynamics is a direct and systematic method to
obtain the equations of conservative motions. This principle
extended to each component of a mixture of conservative fluids is
able to deduce the same number of balance equations than unknown
functions. The method yields a non ambiguous framework for the case
of non-conservative mixtures (with viscosity, diffusion and heat
transfer). Non additional assumption but constitutive behavior
compatible with the second law of thermodynamics is necessary. One
obtains the dynamic Gibbs relation and Fick's law as a consequence
of governing equations.

We have seen that the dependance of an entropy gradient is necessary
for the existence of isentropic waves of acceleration along the
interfaces: the fact that the internal energy depends not only on
the gradient of matter densities but also on the gradient of
entropy, yields a new kind of waves which does not appear in simpler
models. They are exceptional waves in the sense of Boillat and Lax
\cite{Ruggeri} and they appear only in, at least, systems with two
dimensions. These second order waves are of weak energy and
consequently they are not easy to show up. Recent experiments in
space laboratories in micro-gravity conditions, for carbonic dioxide
near its critical point, have showed the possibility of such waves
\cite{Garrabos}. The experimental evidence of such adiabatic waves
with other physical reasons we have presented in the introduction
should strengthen the necessity to take into account of the
dependence of entropy gradients together with density gradients in
the expression of the internal energy for continuum models of
capillarity and phase transitions.\newline

\begin{equation*}
\text{\textbf{Acknowledgements}}
\end{equation*}
This paper was developed during a stay of Tommaso Ruggeri as
visiting professor in L.M.M.T.\ of the University of Aix-Marseille
III and a stay of Henri Gouin as visiting professor in C.I.R.A.M.\
of the University of Bologna with a fellowship of the Italian
GNFM-INDAM and was supported in part (T.R.) by MIUR\ Progetto di
interesse Nazionale \textit{Problemi Matematici Non Lineari di
Propagazione e Stabilit\`{a} nei Modelli del Continuo} Coordinatore
T.\ Ruggeri, by the GNFM-INDAM, and by the Istituto Nazionale di
Fisica Nucleare (INFN).\\
The authors are indebted to the anonymous referees for their
valuable criticism during the review process.

\section{Appendix:}

\subsection{Compatibility conditions for weak discontinuities}

We obtain also the compatibility due to the boundary conditions. For this
aim we rewrite the variations of the Hamilton action when the equations of
the motion (\ref{App. 2}) are verified. Then, from relation (\ref{Hamilton
variations}), we obtain,
\begin{eqnarray*}
&&\delta _{i}I =\int_{t_{1}}^{t_{2}}\int_{\partial D_{t}}\ \ g\ \rho _{i}%
\mathbf{K}_{i}^{T}\ \mathbf{\zeta }_{i} \\
&&-\mathbf{n}.\left( \rho _{i}R_{i}\ \mathbf{\zeta }_{i}-\rho _{i}\mathbf{u}%
_{i}\ \mathbf{K}_{i}^{T}\ \mathbf{\zeta }_{i}-\mathbf{\Phi }_{i}\ \frac{%
\partial \rho _{i}}{\partial \mathbf{x}}\ \mathbf{\zeta }_{i}-\mathbf{\Psi }%
_{i}\ \frac{\partial s_{i}}{\partial \mathbf{x}}\;\mathbf{\zeta }_{i}-\rho
_{i}\mathbf{\Phi }_{i\ }\mathrm{div\ }\mathbf{\zeta }_{i}\right) \ d\sigma
_{x}dt.
\end{eqnarray*}
For a vector field $\mathbf{x\in }$ $D_{t}\ \mathbf{\rightarrow \ \zeta }%
_{i} $ vanishing with its first derivatives on the boundary $\partial D_{t},$
we deduce immediately on a surface of discontinuity $\Sigma _{t}$ (where $%
\rho _{i},$ $s_{i}$ and its first derivatives are continuous and there are
jumps for the second derivative)\ the value of the variation of the Hamilton
action,
\begin{eqnarray*}
&&\delta _{i}I =\int_{t_{1}}^{t_{2}}\int_{\Sigma _{t}}\ \ g\ \left[ \rho _{i}%
\mathbf{K}_{i}^{T}\ \mathbf{\zeta }_{i}\right] \\
&&-\mathbf{n}.\left[ \rho _{i}R_{i}\ \mathbf{\zeta }_{i}-\rho _{i}\mathbf{u}%
_{i}\ \mathbf{K}_{i}^{T}\ \mathbf{\zeta }_{i}-\mathbf{\Phi }_{i}\ \frac{%
\partial \rho _{i}}{\partial \mathbf{x}}\ \mathbf{\zeta }_{i}-\mathbf{\Psi }%
_{i}\ \frac{\partial s_{i}}{\partial \mathbf{x}}\;\mathbf{\zeta }_{i}-\rho
_{i}\mathbf{\Phi }_{i\ }\mathrm{div\ }\mathbf{\zeta }_{i}\right] \ d\sigma
_{x}dt \\
&&\equiv -\int_{t_{1}}^{t_{2}}\int_{\Sigma _{t}}\mathbf{n\ .}\left[ \rho
_{i}R_{i}\right] \ \mathbf{\zeta }_{i}\ d\sigma _{x}dt.
\end{eqnarray*}
\newline
Due to the fact that $\delta _{i}I=0$ for a vector field $\mathbf{x\in }$ $%
D_{t}\ \mathbf{\rightarrow \ \zeta }_{i}\mathbf{,}$ we obtain $\left[ \rho
_{i}R_{i}\right] =0,$ and deduce the compatibility conditions (\ref{sigmai})
across the wave front,
\begin{equation*}
\left[ \mathrm{div\ }\mathbf{\Phi }_{i}\right] \equiv \sum_{j=1}^{2}\left(
C_{ij}\ \delta ^{2}\rho _{j}+D_{ij}\ \delta ^{2}s_{j}\right) =0.
\end{equation*}

\subsection{Mixture with an entropy for the total fluid}

The equations of balance of matter densities $\rho ,$ $\rho _{1}\;$are in
the form
\begin{equation*}
{\frac{\partial \rho }{\partial t}}\ +\ \text{div}(\rho {\mathbf{u}})=0,
\end{equation*}
\begin{equation}
{\frac{\partial \rho _{1}}{\partial t}}\ +\ \text{div}(\rho _{1}{\mathbf{u}}%
_{1})=0,  \label{mass 1}
\end{equation}
\newline
where $\rho $ and ${\mathbf{u}}$ denote the density and the velocity of the
total mixture, $\rho _{1}$ and ${\mathbf{u}}_{1}$ denote the density and the
velocity of one of the two components (it is equivalent to consider the
total density $\rho $ and the concentration$\ c=$ $\rho _{1}/\rho $ betwen
the two components of the mixture such that eq. (\ref{mass 1}) is equivalent
to ${\ \partial }\left( \rho c\right) /\partial t\ +\ $div$(\rho c{\mathbf{u}%
})=0$). For conservative motions, the equation of conservation of the total
specific entropy $s$ is
\begin{equation*}
{\frac{\partial \rho s\ }{\partial t}}\ +\ \mathrm{div}(\rho s{\mathbf{u}}%
)=0.
\end{equation*}
We consider a volume potential energy of the mixture in the form
\begin{equation*}
\varepsilon =\varepsilon (\rho ,\rho _{1},s,s_{1},{\mathbf{\nabla }}\rho ,{%
\mathbf{\nabla }}\rho _{1:},{\mathbf{\nabla }}s,\omega ).
\end{equation*}
with $\omega =\dfrac{1}{2}\,{\mathbf{w}}^{2}$, where ${\mathbf{w}}={\mathbf{u%
}}-{\mathbf{u}}_{1}$. As in the section 2, it is possible\ to deduce the
equations of motions through the Hamilton action. We denote
\begin{equation*}
\rho\, \theta \ =\ \frac{\widehat{\partial }\varepsilon }{\widehat{%
\partial }s\ },~h=\frac{\widehat{\partial }\varepsilon }{\widehat{\partial }%
\rho }\mathrm{,}\ h_{1}=\frac{\widehat{\partial }\varepsilon }{\widehat{%
\partial }\rho _{1}}\ \ \mathrm{and}\ \ a\ =\frac{\partial \varepsilon }{%
\partial \omega }.
\end{equation*}
(In fact, the potential energy of the mixture depends only on the
insentropic invariants $\left( {\mathbf{\nabla }}\rho \right) ^{2},\left( {%
\mathbf{\nabla }}\rho _{1}\right) ^{2},{\mathbf{\nabla }}\rho \cdot {\mathbf{%
\nabla }}\rho _{1},\left( {\mathbf{\nabla }}s\right) ^{2}$). By analogous
calculations associated with a Lagrangian of the mixture in the form

\begin{equation*}
L=\frac{1}{2}\rho \mathbf{u}^{2}-\varepsilon ,
\end{equation*}
and virtual motions such as $\mathbf{X}=\mathbf{\Xi }(t,\mathbf{x,}\varkappa
\mathbf{)}$ and $\mathbf{X}_{1}=\mathbf{\Xi }_{1}(t,\mathbf{x,}\varkappa _{1}%
\mathbf{),\;}$we obtain as in section 2, the equations of motion
\begin{eqnarray}
&&\frac{\partial {\rho \mathbf{u}}}{\partial t}\,+\,\mathrm{div}(\rho {%
\mathbf{u}}\otimes {\mathbf{u}})+  \label{moment global} \\
&&-\left( \frac{\partial }{\partial t}(a{\mathbf{w}})+a\frac{\partial {%
\mathbf{u}}}{\partial {\mathbf{x}}}^{T}{\mathbf{w}}+\mathrm{div}(a\ {\mathbf{%
w}}\otimes {\mathbf{u}})\right) =\rho\, \theta \,{\mathbf{\nabla }}\,s-\rho {%
\mathbf{\nabla }}\,h  \notag
\end{eqnarray}
and
\begin{equation}
\left( \frac{\partial }{\partial t}(a{\mathbf{w}})+a\frac{\partial {\mathbf{u%
}}_{1}}{\partial {\mathbf{x}}}^{T}{\mathbf{w}}+\mathrm{div}(a\ {\mathbf{w}}%
\otimes {\mathbf{u}}_{1})\right) +\rho _{1}{\mathbf{\nabla }}\,h_{1}=0
\label{moment 1}
\end{equation}
with
\begin{equation*}
\rho \,\theta =\rho \epsilon _{,s}-\mathrm{div\ }\mathbf{\Psi ,}\ h\ =\rho
\epsilon _{,\rho }-\mathrm{div\ }\mathbf{\Phi }\mathrm{\ },\mathrm{\ \ and}\
\ \ h_{1}=\rho _{1}\epsilon _{,\rho _{1}}-\mathrm{div\ }\mathbf{\Phi }_{1}%
\mathrm{\ },
\end{equation*}
where
\begin{equation*}
\mathbf{\Psi \ }=\frac{\partial \varepsilon }{\partial {\mathbf{\nabla }}s}%
,\ \mathbf{\Phi \ }=\frac{\partial \varepsilon }{\partial {\mathbf{\nabla }}%
\rho }\ ,\mathbf{\Phi }_{1}=\frac{\partial \varepsilon }{\partial {\mathbf{%
\nabla }}\rho _{1}}.
\end{equation*}
By summing eqs (\ref{moment global},\ref{moment 1}), we obtain the balance
equation for the total momentum in a divergence form
\begin{equation*}
\frac{\partial {\rho \mathbf{u}}}{\partial t}\,+\,\mathrm{div}\left( \rho \ {%
\mathbf{u}}\otimes {\mathbf{u}}-\rho a\ {\mathbf{w}}\otimes {\mathbf{w}}-%
\mathbf{\sigma }\right) =0
\end{equation*}
where$\ \ \mathbf{\sigma }\ \ $is the stress tensor such that
\begin{equation*}
\sigma _{\nu \gamma }=(-P+\rho \ \mathrm{div\ }\mathbf{\Phi +}\rho _{1}\
\mathrm{div\ }\mathbf{\Phi }_{1})~\delta _{\nu \gamma }-\Phi _{\nu }\rho
_{,\gamma }-\Phi _{1\nu }\rho _{1,\gamma }~-\Psi _{\nu }s_{,\gamma },
\end{equation*}
where $P=\epsilon -\rho \epsilon _{,\rho }-\rho _{1}\epsilon _{,\rho _{1}}$.$%
\ $The equation of energy of the total mixture is also obtained in the
divergence form
\begin{equation*}
\frac{\partial }{\partial t}\left( \frac{1}{2}\rho \mathbf{u}%
^{2}+\varepsilon -a\ \mathbf{w}^{2}\right) +\mathrm{div}\left( \left( \frac{1%
}{2}\rho \mathbf{u}^{2}-a\ \mathbf{w.u}-\sigma \right) \mathbf{u}\ -\mathbf{U%
}+\varepsilon \ \mathbf{u}\right) =0.
\end{equation*}
with $\mathbf{U}=\dfrac{d\rho }{dt}\;\mathbf{\Phi \ +\ }\dfrac{d_{1}\rho _{1}%
}{dt}\;\mathbf{\Phi }_{1}+\dfrac{ds}{dt}\;\mathbf{\Psi }$ and it is possible
by a similar calculation than in section 3 to deduce an acceleration wave
associated with the entropy gradient.

\end{document}